\documentclass[11pt]{amsart}
\usepackage{amssymb}
\usepackage{amsfonts}
\usepackage{indentfirst}
\usepackage{amsopn}
\usepackage{amsmath}
\usepackage{amsthm}
\usepackage{amscd}
\usepackage{graphicx}
\usepackage{epstopdf}

\makeatletter\@addtoreset {equation}{section}\makeatother

\begin{document}

\title[Rogue waves on the periodic standing waves]{\bf Periodic standing waves \\
in the focusing nonlinear Schr\"{o}dinger equation: \\
rogue waves and modulation instability}

\author{Jinbing Chen}
\address[J. Chen]{School of Mathematics, Southeast University, Nanjing, Jiangsu 210096, P.R. China}
\email{cjb@seu.edu.cn}

\author{Dmitry E. Pelinovsky}
\address[D.E. Pelinovsky]{Department of Mathematics, McMaster University, Hamilton, Ontario, Canada, L8S 4K1}
\email{dmpeli@math.mcmaster.ca}
\address[D.E. Pelinovsky]{Institute of Applied Physics RAS, Nizhny Novgorod, 603950, Russia}

\author{Robert E. White}
\address[R.E. White]{Department of Mathematics, McMaster University, Hamilton, Ontario, Canada, L8S 4K1}
\email{whitere@mcmaster.ca}

\date{\today}
\maketitle

\begin{abstract}
We present exact solutions for rogue waves arising on the background of
periodic standing waves in the focusing nonlinear Schr\"{o}dinger equation. The exact solutions are
obtained by characterizing the Lax spectrum related to the periodic standing waves and by using
the one-fold Darboux transformation. The magnification factor of the rogue waves is
computed in the closed analytical form. We relate the rogue wave solutions to
the modulation instability of the background of the periodic standing waves.
\end{abstract}

\section{Introduction}

We address rogue waves described by the focusing nonlinear Schr\"{o}dinger (NLS) equation:
\begin{equation}
i \psi_t + \frac{1}{2} \psi_{xx} + |\psi|^2 \psi = 0.
\label{nls}
\end{equation}
The {\em canonical rogue wave} was derived in \cite{Akh85,Peregrine} in the exact form:
\begin{equation}
\label{rogue-basic}
\psi(x,t) = \left[ 1 - \frac{4 (1+2it)}{1 + 4 x^2 + 4 t^2} \right] e^{it}.
\end{equation}
As $|t| + |x| \to \infty$, the rogue wave (\ref{rogue-basic}) approaches the constant-amplitude wave
$\psi_0(t) = e^{it}$, which is modulationally unstable \cite{ZakOst}. The rogue wave reaches its
maximum amplitude $|\psi(0,0)| = 3$ at $(x,t) = (0,0)$ on the background $|\psi_0(t)| = 1$, hence it
achieves a triple magnification over the constant-amplitude wave.
Other rational solutions for rogue waves in the NLS equation
were constructed by using Darboux transformations in \cite{Akh,Matveev,JYang}.
Further connection between rogue waves and the modulationally unstable constant-amplitude waves
was investigated by using the inverse scattering method \cite{Bil1,Bil2,Bil3},
asymptotic analysis \cite{Biondini}, and the finite-gap theory \cite{GS1,GS2}.

{\em Periodic standing waves} in the focusing NLS equation take the form
\begin{equation}
\label{standing-wave}
\psi(x,t) = U(x) e^{2ibt},
\end{equation}
where $U$ is a $L$-periodic complex-valued function and $2 b$ is a real frequency parameter.
These periodic standing waves are known to be modulationally unstable with respect to long-wave perturbations \cite{BHJ,DS}
(see also recent studies in \cite{DU,DU2}). Rogue waves have been
observed numerically on the background of the modulationally unstable periodic standing waves \cite{AZ1,AZ2}.
Construction of solutions of the NLS equation (\ref{nls}) for
such rogue waves on the background of the periodic standing waves was first performed numerically in \cite{CalSch,Kedziora}
by applying the one-fold Darboux transformation to the numerically constructed solutions
of the Lax equations. It was only recently in \cite{CPkdv,CPnls} that the exact solutions
for such rogue waves were constructed in the closed form for the ${\rm dn}$-periodic and
${\rm cn}$-periodic elliptic waves. A more complete analysis of rogue waves on the background of periodic elliptic waves
was developed in \cite{Feng}.

Rogue waves on the multi-phase solutions and their magnification factors were studied in \cite{Tovbis1,Tovbis2}
by using Riemann Theta functions. In the recent study \cite{CPW}, we have also studied numerically
the rogue waves arising on the double-periodic (both in space and in time) solutions
to the focusing NLS equation (\ref{nls}).

Let us now give a mathematical definition of a rogue wave on the background of a periodic standing wave.
Let $\psi_0(x,t)$ be a periodic standing wave of the focusing NLS equation (\ref{nls}) in the form (\ref{standing-wave}).
We say that $\psi(x,t)$ is {\em a rogue wave on the background of the periodic standing wave} if it satisfies
\begin{equation}
\label{rogue-wave-def}
\inf_{x_0,t_0,\alpha_0 \in \mathbb{R}} \sup_{x \in \mathbb{R}} \left| \psi(x,t) - \psi_0(x-x_0,t-t_0) e^{i \alpha_0} \right| \to 0
\quad \mbox{\rm as} \quad t \to \pm \infty.
\end{equation}
The magnification factor of the rogue wave can be defined as a ratio of the maximum of $|\psi|$ to the maximum of $|\psi_0|$ by
\begin{equation}
\label{factor-magnification}
M := \frac{\max_{(x,t) \in \mathbb{R}^2} |\psi(x,t)|}{\max_{(x,t) \in \mathbb{R}^2} |\psi_0(x,t)|}.
\end{equation}
Although the physical definition of the magnification factor is the ratio of the maximum of $|\psi|$ to the mean value of $|\psi_0|$
\cite{Tovbis1,CPW}, which is bigger than $M$ in (\ref{factor-magnification}), we shall use the definition (\ref{factor-magnification})
for computations of the magnification factor $M$ in the closed analytical form.

The purpose of this work is to construct the exact solutions for rogue waves on the background
of the periodic standing waves. We improve the previous work \cite{CPnls} in several ways.

First, we develop a general scheme for integrability
of the Hamiltonian system which arises in the algebraic method with one eigenvalue \cite{ZhouJMP,ZhouStudies}
(nonlinearization of the linear equations was pioneered in \cite{Cao1}). This scheme allows us to integrate
the standing wave reduction of the focusing NLS equation (\ref{nls}) and to relate parameters
of the periodic standing waves to eigenvalues in the Lax spectrum. This characterization of
eigenvalues at the end points of the Lax spectrum complements the resolvent method developed
for the same purpose in \cite{Kam} (see also review in \cite{Kam-review}).

Second, we introduce a new representation
for the second growing solution to the linear equations that is applicable to
every admissible eigenvalue of the Lax spectrum. The previous representation
of the second growing solution in \cite{CPnls} only works for one eigenvalue
in the Lax spectrum of the dn-periodic wave but is singular for the other eigenvalue.
The new representation also gives a simpler expression to control the linear growth
of the second growing solution everywhere on the $(x,t)$ plane.

Third, we compute the explicit expression of the magnification factor $M$ in (\ref{factor-magnification})
for a general elliptic wave solution with nontrivial phase. This analytical expression
generalizes the previous computations of the magnification factors of the constant-amplitude wave,
the ${\rm dn}$-periodic wave, and the ${\rm cn}$-periodic wave.

Fourth, we relate the existence of such rogue wave solutions to the modulation instability of the periodic wave background
studied in \cite{DS}. We obtain an exceptional curve on the two-parameter plane of the general elliptic wave solutions,
at which our new solution associated with one eigenvalue in the Lax spectrum is not fully localized
on the $(x,t)$ plane and violates the definition (\ref{rogue-wave-def}) of the rogue wave on the background of the
periodic standing wave. Instead, our solution describes an algebraic soliton traveling on the background
of the periodic standing wave. This solution is similar to the solutions obtained on the ${\rm dn}$-periodic wave
in the mKdV equation \cite{CPkdv}. We show that the exceptional curve coincides with the one obtained
in \cite{DS} from the condition that the modulation instability band in the linearization
of the focusing NLS equation (\ref{nls}) at the periodic standing wave is tangential to the imaginary axis
at the origin of the complex plane.

Fifth, we visualize numerically the Lax spectrum, the admissible eigenvalues, the modulation instability bands,
and the rogue waves.

The article is organized as follows. Section 2 describes the algebraic method with one eigenvalue
which is used to relate the periodic standing waves with the eigenvalues in their Lax spectrum.
Section 3 characterizes parameters of the periodic standing waves and their representation in terms
of the Jacobian elliptic functions. Section 4 characterizes the periodic and linearly growing solutions of the Lax equations at
eigenvalues of the Lax spectrum related to the periodic standing waves. Section 5 presents
exact expressions for the rogue waves and for their magnification factors. Section 6
discusses the relation between the existence of such rogue waves and the modulation instability
bands for the periodic standing waves.  Appendix A gives details of numerical computations
of the Lax spectrum for the periodic standing waves. Appendix B contains technical details
of computations involving Jacobian elliptic functions.

\section{Algebraic method with one eigenvalue}

The NLS equation (\ref{nls}) with $\psi \equiv u$ appears as a compatibility condition $\varphi_{xt} = \varphi_{tx}$
of the following pair of linear equations on $\varphi \in \mathbb{C}^2$:
\begin{equation}\label{3.1}
\varphi_x = Q(\lambda,u) \varphi,\qquad \qquad
Q(\lambda,u) = \left(\begin{array}{cc} \lambda & u \\ -\bar{u} & -\lambda \end{array} \right)
\end{equation}
and
\begin{equation}\label{3.2}
\varphi_t = S(\lambda,u) \varphi,\qquad
S(\lambda,u) = i \left(\begin{array}{cc}
\lambda^2 + \frac{1}{2} |u|^2 & \frac{1}{2} u_x + \lambda u\\
\frac{1}{2} \bar{u}_x - \lambda\bar{u} & -\lambda^2 - \frac{1}{2} |u|^2\\
\end{array}
\right),
\end{equation}
where $\bar{u}$ is the conjugate of $u$ and $\lambda \in \mathbb{C}$ is a spectral parameter.

The procedure of computing a new solution $\psi \equiv \hat{u}$
of the NLS equation (\ref{nls}) from another solution $\psi \equiv u$ is well-known \cite{Akh,CPnls,Feng}.
Let $\varphi = (p_1,q_1)^t$ be any nonzero solution to the linear equations (\ref{3.1}) and (\ref{3.2})
for a fixed value $\lambda = \lambda_1$. The new solution is given by the one-fold Darboux transformation
\begin{equation}
\label{1-fold}
\hat{u} = u + \frac{2 (\lambda_1 + \bar{\lambda}_1) p_1 \bar{q}_1}{|p_1|^2 + |q_1|^2},
\end{equation}
and in particular, it may represent a rogue wave $\hat{u}$ on the background $u$ satisfying the limits (\ref{rogue-wave-def}).
The main question is which value $\lambda_1$ to fix and which nonzero solution $\varphi$ to the linear equations
(\ref{3.1})--(\ref{3.2}) to take. It is clear from (\ref{1-fold}) that $\lambda_1 \notin i \mathbb{R}$, because if
$\lambda_1 + \bar{\lambda}_1 = 0$, then $\hat{u} = u$.
For the periodic standing wave in the form (\ref{standing-wave}), we show that the admissible values of $\lambda_1$ are defined by
the algebraic method with one eigenvalue, which is a particular case of a general method of nonlinearization
of the linear equations on $\varphi$ developed in \cite{Cao1} and \cite{ZhouJMP,ZhouStudies}.

Let $\varphi = (p_1,q_1)^t$ be a nonzero solution of the linear equations (\ref{3.1})--(\ref{3.2}) with
fixed $\lambda = \lambda_1$ such that $\lambda_1 + \bar{\lambda}_1 \neq 0$. We introduce the following
relation between the solution $u$ to the NLS equation (\ref{nls}) and the squared eigenfunction
$\varphi = (p_1,q_1)^t$:
\begin{equation}\label{3.3}
u = p_1^2 + \bar{q}_1^2.
\end{equation}
We intend to determine the value of $\lambda_1$ for which the relation (\ref{3.3}) holds
in $(x,t)$.

Assuming (\ref{3.3}), the spectral problem (\ref{3.1}) is nonlinearized into the following Hamiltonian system:
\begin{equation}\label{3.4}
\frac{d p_1}{d x} = \frac{\partial H}{\partial q_1}, \quad
\frac{d q_1}{d x} = - \frac{\partial H}{\partial p_1},
\end{equation}
where the Hamiltonian function is
\begin{equation}\label{3.5}
H = \lambda_1 p_1 q_1 + \bar{\lambda}_1 \bar{p}_1 \bar{q}_1 + \frac{1}{2} |p_1^2 + \bar{q}_1^2|^2
\end{equation}
and the ordinary derivatives in $x$ (at fixed $t$) are intentionally used in (\ref{3.4})
to emphasize that the Hamiltonian system in $x$ (at fixed $t$) is of degree two.
The value of $H$ is a constant of motion for the Hamiltonian system
(\ref{3.4}) together with another constant of motion:
\begin{eqnarray}
F = i (p_1 q_1 - \bar{p}_1 \bar{q}_1).
\label{3.9}
\end{eqnarray}
Hence, the Hamiltonian system of degree two is Liouville integrable.

Assuming (\ref{3.3}), the time-evolution problem (\ref{3.2}) is nonlinearized into another Hamiltonian system given by
\begin{equation}\label{a1}
\frac{d p_1}{dt} = \frac{\partial K}{\partial q_1}, \quad
\frac{d q_1}{dt} = -\frac{\partial K}{\partial p_1},
\end{equation}
where the Hamiltonian function is
\begin{eqnarray}
\nonumber
K & = & i \left[
2\lambda_1^2p_1q_1-2\bar{\lambda}_1^2\bar{p}_1\bar{q}_1+|p_1^2+\bar{q}_1^2|^2(p_1q_1-\bar{p}_1\bar{q}_1) \right. \\
\label{a2}
& \phantom{t} & \left.
+ (\lambda_1p_1^2-\bar{\lambda}_1\bar{q}_1^2)(\bar{p}_1^2+q_1^2)+(p_1^2+\bar{q}_1^2)(\lambda_1q_1^2-\bar{\lambda}_1\bar{p}_1^2) \right]
\end{eqnarray}
and the ordinary derivatives in $t$ (at fixed $x$) are intentionally used in (\ref{a1})
to emphasize that the Hamiltonian system in $t$ (at fixed $x$) is of degree two.
The Hamiltonian system (\ref{a1}) is also Liouville integrable because both $K$ in (\ref{a2})
and $F$ in (\ref{3.9}) are constant in $t$, where the latter conservation
can be verified directly as follows:
\begin{eqnarray*}
\frac{d F}{dt} & = & - q_1 \left[ (2\lambda_1^2 + |u|^2) p_1 + (u_x + 2 \lambda_1 u) q_1 \right]
 - p_1 \left[ -(2\lambda_1^2+|u|^2) q_1 + (\bar{u}_x - 2 \lambda_1 \bar{u}) p_1 \right] \\
& \phantom{t} & - \bar{q}_1 \left[ (2\bar{\lambda}_1^2 + |u|^2) \bar{p}_1 + (\bar{u}_x + 2 \bar{\lambda}_1 \bar{u}) \bar{q}_1 \right]
- \bar{p}_1 \left[ -(2\bar{\lambda}_1^2+|u|^2) \bar{q}_1 + (u_x - 2 \bar{\lambda}_1 u) \bar{p}_1 \right] \\
& = & - \left[ \bar{u} u_x + u \bar{u}_x + 2 u (\lambda_1 q_1^2 - \bar{\lambda}_1 \bar{p}_1^2) + 2 \bar{u} (\bar{\lambda}_1 \bar{q}_1^2 - \lambda_1^2 p_1^2) \right] \\ &=& 0,
\end{eqnarray*}
thanks to the constraint that follows from differentiating (\ref{3.3}) in $x$ and substituting (\ref{3.4}):
\begin{eqnarray}
\label{4a}
u_x = 2 (\lambda_1 p_1^2 - \bar{\lambda}_1 \bar{q}_1^2) + 2 (p_1^2 + \bar{q}_1^2) (p_1 q_1 - \bar{p}_1 \bar{q}_1).
\end{eqnarray}
Similarly, we can check that the Hamiltonian systems (\ref{3.4}) and (\ref{a1})
commute in the sense of $\{ H, K \} = 0$, where the Poisson bracket is given by
$$
\{ f, g \} := \frac{\partial f}{\partial p_1}\frac{\partial g}{\partial q_1}-\frac{\partial f}{\partial q_1}\frac{\partial g}{\partial p_1}
+\frac{\partial f}{\partial \bar{p}_1}\frac{\partial g}{\partial \bar{q}_1}-\frac{\partial f}{\partial \bar{q}_1}\frac{\partial g}{\partial \bar{p}_1}.
$$
It follows from (\ref{3.5}), (\ref{a2}), and (\ref{4a}) that
\begin{eqnarray*}
\{H,K\} &=& i \left[ (\lambda_1q_1 + \bar{u}p_1)((2\lambda_1^2+|u|^2)p_1+(u_x+2\lambda_1u)q_1) \right. \\&&
\left. +(\lambda_1p_1+uq_1)((\bar{u}_x-2\lambda_1\bar{u})p_1-(2\lambda_1^2+|u|^2)q_1) \right. \\&&
\left. -(\bar{\lambda}_1\bar{q}_1 + u\bar{p}_1)((2\bar{\lambda}_1^2+|u|^2)\bar{p}_1+(\bar{u}_x+2\bar{\lambda}_1\bar{u})\bar{q}_1) \right. \\&&
\left. -(\bar{\lambda}_1\bar{p}_1+\bar{u}\bar{q}_1)((u_x-2\bar{\lambda}_1u)\bar{p}_1-(2\bar{\lambda}_1^2+|u|^2)\bar{q}_1) \right]\\
&=& i \left[ \lambda_1(u_x q_1^2+\bar{u}_x p_1^2)-\bar{\lambda}_1(\bar{u}_x \bar{q}_1^2 + u_x \bar{p}_1^2)
+(\bar{u}u_x + \bar{u}_x u) (p_1 q_1 - \bar{p}_1 \bar{q}_1) \right] \\
&=& i \left[ u_x (\lambda_1 q_1^2 - \bar{\lambda}_1 \bar{p}_1^2 + \bar{u}(p_1 q_1 - \bar{p}_1 \bar{q}_1)) +
\bar{u}_x (\lambda_1 p_1^2 - \bar{\lambda}_1 \bar{q}_1^2 + u (p_1 q_1 - \bar{p}_1 \bar{q}_1)) \right] \\
&=& 0.
\end{eqnarray*}
Since $H$ and $K$ commute, then $H$ is constant in $t$ and $K$ is constant in $x$.

We claim and show next that the constraint (\ref{3.3}) can only represent
the class of standing and traveling wave solutions of the NLS equation (\ref{nls}) in the form
\begin{equation}
\label{travelling-wave}
\psi(x,t) \equiv u(x,t) = U(x+ct) e^{2 i b t},
\end{equation}
where $U$ is a complex-valued function and $(c,b)$ are real-valued parameters. For
the standing and traveling wave solution of the form (\ref{travelling-wave}),
the constraint (\ref{3.3}) determines solutions of the
Hamiltonian systems (\ref{3.4}) and (\ref{a1}) in the form
\begin{equation}
\label{travelling-wave-p-q}
p_1(x,t) = P_1(x+ct) e^{i b t}, \quad q_1(x,t) = Q_1(x+ct) e^{-ibt},
\end{equation}
where $P_1$ and $Q_1$ are complex-valued functions. Thanks to the commutativity
of the Hamiltonian systems (\ref{3.4}) and (\ref{a1}), this implies that the time evolution
of $u(x,t)$, $p_1(x,t)$, and $q_1(x,t)$ in $t$ is trivially given by (\ref{travelling-wave}) and (\ref{travelling-wave-p-q}),
whereas the functional dependence of $u(x,t)$, $p_1(x,t)$, and $q_1(x,t)$ on $x$ is non-trivial.
Due to this reason, we abuse the notations and only refer to the $x$-dependence
of $u$, $p_1$, and $q_1$ in what follows before Section 4. In particular, we rewrite
(\ref{4a}) in the equivalent form:
\begin{equation}\label{3.13}
\frac{d u}{d x} + 2 i F u = 2 (\lambda_1 p_1^2 - \bar{\lambda}_1 \bar{q}_1^2),
\end{equation}
where the ordinary derivative in $x$ is now used. By differentiating (\ref{3.13})
and using (\ref{3.4}), (\ref{3.5}), and (\ref{3.9}), we obtain
\begin{equation}\label{3.14}
\frac{d^2 u}{d x^2} + 2 |u|^2 u + 2 i F \frac{du}{dx} - 4 H u
= 4 (\lambda_1^2 p_1^2 + \bar{\lambda}_1^2 \bar{q}_1^2).
\end{equation}
Eliminating the squared eigenfunctions $p_1^2$ and $\bar{q}_1^2$ from (\ref{3.3}) and
(\ref{3.13}) and substituting the result into (\ref{3.14}) yields a closed second-order equation:
\begin{equation}
\label{LN-2}
\frac{d^2 u}{dx^2} + 2 |u|^2 u + 2 i c \frac{du}{dx} - 4b u = 0,
\end{equation}
where $c$ and $b$ are real parameters given by
\begin{equation}
\label{3.20a}
\left\{
\begin{array}{l}
c = F + i(\lambda_1 - \bar{\lambda}_1), \\
b = H + i F(\lambda_1 - \bar{\lambda}_1) + |\lambda_1|^2.
\end{array} \right.
\end{equation}
The differential equation (\ref{LN-2}) is a standing and traveling wave reduction
of the NLS equation (\ref{nls}) for the solutions in the form (\ref{travelling-wave}) with
$u \equiv U$. Hence, the constraint (\ref{3.3}) can only represent
the class of standing and traveling wave solutions of the NLS equation (\ref{nls}).

In order to complete the algebraic method with one eigenvalue and to integrate
the second-order equation (\ref{LN-2}), we represent the Hamiltonian system (\ref{3.4})
by using the following Lax equation:
\begin{equation}\label{3.7}
\frac{d}{dx} W(\lambda) = [Q(\lambda,u),W(\lambda)],
\end{equation}
where $Q(\lambda,u)$ is given by (\ref{3.1}), $u$ is given by (\ref{3.3}),
and $W(\lambda)$ is represented in the form:
\begin{equation}\label{3.6}
\displaystyle
W(\lambda) =\left(\begin{array}{cc}
W_{11}(\lambda) & W_{12}(\lambda) \\
\overline{W}_{12}(-\lambda) & -\overline{W}_{11}(-\lambda) \end{array}\right),
\end{equation}
with the entries
\begin{equation}
\label{W-11-12}
W_{11}(\lambda) = 1 - \left(\frac{p_1 q_1}{\lambda-\lambda_1}-\frac{\bar{p}_1\bar{q}_1}{\lambda+\bar{\lambda}_1}\right), \qquad
W_{12}(\lambda) = \frac{p_1^2}{\lambda-\lambda_1}+\frac{\bar{q}^2_1}{\lambda+\bar{\lambda}_1}.
\end{equation}
By using (\ref{3.3}), (\ref{3.5}), (\ref{3.9}), and (\ref{3.13}),
we can rewrite the pole decompositions (\ref{W-11-12}) in the form:
\begin{equation}\label{3.19}
W_{11}(\lambda) = \frac{\lambda^2 + i c \lambda + \frac{1}{2} |u|^2 - b}{(\lambda - \lambda_1)(\lambda + \bar{\lambda}_1)}, \qquad
W_{12}(\lambda) = \frac{u \lambda + \frac{1}{2} \frac{du}{dx} + i c u}{(\lambda - \lambda_1)(\lambda + \bar{\lambda}_1)}.
\end{equation}
It follows from (\ref{3.6}) and (\ref{W-11-12}) that
\begin{eqnarray}
\label{det-W}
\det W(\lambda) = - \left[ W_{11}(\lambda) \right]^2 - W_{12}(\lambda) \bar{W}_{12}(-\lambda)
\end{eqnarray}
can be reduced with the help of (\ref{3.5}) and (\ref{3.9}) to the form
\begin{eqnarray}
\label{det-W-1}
\det W(\lambda) = - 1 + \frac{2 \left[ H - i F (\lambda - \lambda_1 + \bar{\lambda}_1)\right]}{(\lambda - \lambda_1) (\lambda + \bar{\lambda}_1)}.
\end{eqnarray}
This expression proves that ${\rm det} W(\lambda)$ contains only simple poles at $\lambda_1$ and $-\bar{\lambda}_1$ and
that ${\rm det} W(\lambda)$ is constant in $(x,t)$. Moreover,
if $\lambda_1 + \bar{\lambda}_1 \neq 0$ and $(H,F) \neq (0,0)$,
then the residue terms at $\lambda_1$ and $-\bar{\lambda}_1$ are nonzero.

The $(1,2)$-component of the Lax equation (\ref{3.7}) with (\ref{3.6}) and (\ref{3.19}) is equivalent to
the second-order equation (\ref{LN-2}). On the other hand, the first-order invariants for the second-order equation (\ref{LN-2})
follow from the properties of ${\rm det} W(\lambda)$. By using (\ref{3.6}) and (\ref{3.19}), we obtain
\begin{eqnarray}
\label{det-W-2}
\det W(\lambda) = - \frac{P(\lambda)}{(\lambda-\lambda_1)^2 (\lambda+\bar{\lambda}_1)^2},
\end{eqnarray}
where
\begin{equation}\label{3.24}
P(\lambda) := \left( \lambda^2 + i c \lambda + \frac{1}{2} |u|^2 - b \right)^2 -
\left( u \lambda + \frac{1}{2}  \frac{du}{dx} + i cu \right) \left( \bar{u} \lambda - \frac{1}{2}  \frac{d \bar{u}}{dx} + i c \bar{u} \right).
\end{equation}
Since $P(\lambda)$ is independent of $(x,t)$, expanding the polynomial $P(\lambda)$ in powers of $\lambda$ gives two
first-order invariants for the second-order equation (\ref{LN-2}) in the form:
\begin{equation}
\label{Constant-2b}
\left|  \frac{du}{dx} \right|^2 + |u|^4 - 4 b |u|^2 = 8 d
\end{equation}
and
\begin{equation}
\label{Constant-2a}
i \left( \frac{du}{dx}\bar{u} - u  \frac{d \bar{u}}{dx} \right) - 2 c |u|^2 = 4 a
\end{equation}
where $d$ and $a$ are real parameters appearing in the coefficients of the polynomial $P(\lambda)$ given by
\begin{equation}
\label{Polynomial-2}
P(\lambda) = \lambda^4 + 2 i c \lambda^3 - (c^2 + 2b) \lambda^2 + 2i (a - bc) \lambda + b^2 - 2 ac + 2 d.
\end{equation}
If $\lambda_1$ is a root of $P(\lambda)$, so is $-\bar{\lambda}_1$, thanks to the symmetry of the coefficients
in $P(\lambda)$. The admissible values for $\lambda_1$ are given by four roots of $P(\lambda)$ which are
symmetric about the imaginary axis. If $\lambda_1$ is a simple root of $P(\lambda)$, then ${\rm det} W(\lambda)$ has simple poles
at $\lambda_1$ and $-\bar{\lambda}_1$ in the quotient (\ref{det-W}). Since the residue terms in (\ref{det-W}) are nonzero
if $\lambda_1 + \bar{\lambda}_1 \neq 0$ and $(H,F) \neq (0,0)$,
it is clear that $\lambda_1$ cannot be a double root of $P(\lambda)$.

\section{Classification of periodic standing waves}
\label{sec-waves}

Here we solve the second-order equation (\ref{LN-2}) by using Jacobian elliptic
functions. The algebraic method with one eigenvalue allows us to define the admissible values of $\lambda_1$
for the periodic standing waves of the NLS equation (\ref{nls}).

First, we note that the transformation
\begin{equation}
\label{transformation-Lorenz}
u(x) = \tilde{u}(x) e^{-icx}, \quad b = \tilde{b} + \frac{1}{4} c^2, \quad d = \tilde{d} + \frac{1}{2} a c,
\quad a = \tilde{a}
\end{equation}
leaves the system (\ref{LN-2}), (\ref{Constant-2b}) and (\ref{Constant-2a}) invariant for tilde variables
and eliminates the parameter $c$. Similarly, $P(\lambda)$ in (\ref{Polynomial-2}) can be written as
\begin{equation}
\label{transformation-Lorenz-P}
P(\lambda) = \left( \lambda + \frac{i}{2} c \right)^4 - 2 \tilde{b} \left( \lambda + \frac{i}{2} c \right)^2
+ 2 i \tilde{a} \left( \lambda + \frac{i}{2} c \right) + \tilde{b}^2 + 2 \tilde{d},
\end{equation}
which shows that the dependence of $P(\lambda)$ on $c$ can be scaled out by a translation in the spectral plane of $\lambda$.
Hence, one can set $c = 0$ without loss of generality. This corresponds to the Lorentz transformation
which generates a standing and traveling wave solution of the NLS equation (\ref{nls}) in the form (\ref{travelling-wave})
from every standing wave solution in the form (\ref{standing-wave}).

The periodic standing waves are divided into two groups
depending on whether the phase of the complex-valued function $u(x)$ is trivial or non-trivial.

\subsection{Periodic standing waves with trivial phase}

Let us set $c = 0$ and $a = 0$, where the first choice is made without loss of generality thanks
to the transformations (\ref{transformation-Lorenz}) and (\ref{transformation-Lorenz-P}),
whereas the second choice trivializes the phase and simplifies the construction of periodic standing waves
in Jacobian elliptic functions. It follows from (\ref{Constant-2a}) with $c = a = 0$ that
$$
\frac{d}{dx} \log\left( \frac{u}{\bar{u}}\right) = 0 \quad \Rightarrow \quad \frac{u}{\bar{u}} = e^{2i\theta},
$$
where real $\theta$ is constant in $x$. Thanks to the rotational symmetry of solutions to the NLS
equation (\ref{nls}), one can take $u(x)$ as a real
function. The first-order quadrature (\ref{Constant-2b}) is rewritten for real $u(x)$ as follows:
\begin{equation}
\label{quad-1}
\left( \frac{du}{dx} \right)^2 + V(u) = 0, \quad V(u) := u^4 - 4 b u^2 - 8d.
\end{equation}
The four roots of $V(u)$ are symmetric and can be enumerated as $\pm u_1$ and $\pm u_2$.
If all four roots are real, one can order them as
\begin{equation}
\label{roots-0}
-u_1 \leq -u_2 \leq 0 \leq u_2 \leq u_1.
\end{equation}
Solutions of the first-order quadrature (\ref{quad-1}) exist either in $[-u_1,-u_2]$ or in $[u_2,u_1]$.
The two solutions are related by the transformation $u \mapsto -u$, hence
without loss of generality, one can consider the positive solution in $[u_2,u_1]$.
This solution is given by the explicit formula
\begin{equation}
\label{dn-solution}
u(x) = u_1 {\rm dn}(u_1 x; k), \quad k := \frac{\sqrt{u_1^2 - u_2^2}}{u_1}.
\end{equation}
If two roots of $V(u)$ are real (e.g., $\pm u_1$) and two roots are complex (e.g., $\pm u_2 = \pm i \nu_2$),
the solution in $[-u_1,u_1]$ is given by the explicit formula
\begin{equation}
\label{cn-solution}
u(x) = u_1 {\rm cn}(\alpha x; k), \quad \alpha := \sqrt{u_1^2 + \nu_2^2}, \quad k := \frac{u_1}{\sqrt{u_1^2 + \nu_2^2}}.
\end{equation}
The connection formulas of roots $\pm u_1$, $\pm u_2$ with the parameters $(b,d)$ in (\ref{quad-1})
are given by
\begin{equation}
\label{roots-0-connection}
\left\{ \begin{array}{l}
4b = u_1^2 + u_2^2, \\
8d = -u_1^2 u_2^2.
\end{array} \right.
\end{equation}
Substituting (\ref{roots-0-connection}) into $P(\lambda)$ given by (\ref{Polynomial-2}) with $c = a = 0$ yields
\begin{equation}
\label{P-poly-0}
P(\lambda) = \lambda^4 - \frac{1}{2} (u_1^2 + u_2^2) \lambda^2 + \frac{1}{16} (u_1^2 - u_2^2)^2.
\end{equation}
The two admissible pairs of eigenvalues are given by
\begin{equation}
\label{eig-dn-0}
\lambda_1^{\pm} = \pm \frac{u_1 + u_2}{2}, \quad
\lambda_2^{\pm} = \pm \frac{u_1 - u_2}{2},
\end{equation}
where $u_1$ is real and $u_2$ is either real for (\ref{dn-solution}) or purely imaginary for (\ref{cn-solution}).

Both waves with the trivial phase can be simplified by using the scaling transformation
of solutions to the NLS equation (\ref{nls}):
\begin{equation}
\label{scaling-transform}
\mbox{\rm if} \; u(x,t) \;\; \mbox{\rm is a solution, so is} \;\; a u(ax,a^2t),
\;\; \mbox{\rm for every} \;\; a \in \mathbb{R}.
\end{equation}
The ${\rm dn}$-periodic wave (\ref{dn-solution})
with $u_1 = 1$ and $u_2 = \sqrt{1 - k^2}$ becomes
\begin{equation}
\label{red-1}
u(x) = {\rm dn}(x;k),  \quad k \in (0,1)
\end{equation}
and the two eigenvalue pairs in (\ref{eig-dn-0}) are real-valued:
\begin{equation}
\label{eig-dn}
\lambda_1^{\pm} = \pm \frac{1}{2} ( 1 + \sqrt{1-k^2}), \quad
\lambda_2^{\pm} = \pm \frac{1}{2} ( 1 - \sqrt{1-k^2}).
\end{equation}
The ${\rm cn}$-periodic wave (\ref{cn-solution}) with $u_1 = k$ and $\nu_2 = \sqrt{1 - k^2}$ becomes
\begin{equation}
\label{red-2}
u(x) = k \; {\rm cn}(x;k), \quad k \in (0,1)
\end{equation}
and the two eigenvalue pairs in (\ref{eig-dn-0}) form a complex quadruplet:
\begin{equation}
\label{eig-cn}
\lambda_1^{\pm} = \frac{1}{2} ( \pm k + i \sqrt{1-k^2}), \quad
\lambda_2^{\pm} =  \frac{1}{2} ( \pm k - i \sqrt{1-k^2}).
\end{equation}
These two cases agree with the outcomes of the algebraic method in \cite{CPnls}
and with the explicit expressions obtained in \cite{Kam}.

\begin{figure}[h!]
	\centering
	\includegraphics[scale=0.5]{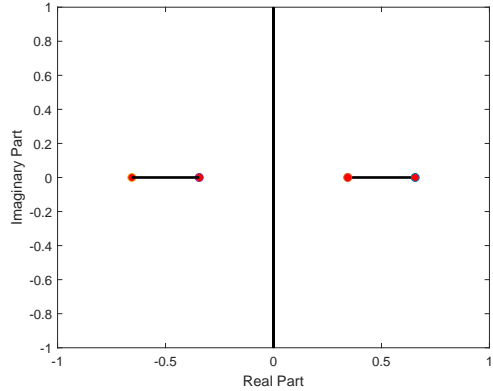}
	\caption{Lax spectrum for the ${\rm dn}$-periodic wave (\ref{red-1}) with $k=0.95$. Red dots
	represent eigenvalues (\ref{eig-dn}).}
	\label{f1}
\end{figure}

\begin{figure}[h!]
	\centering
	\includegraphics[scale=0.5]{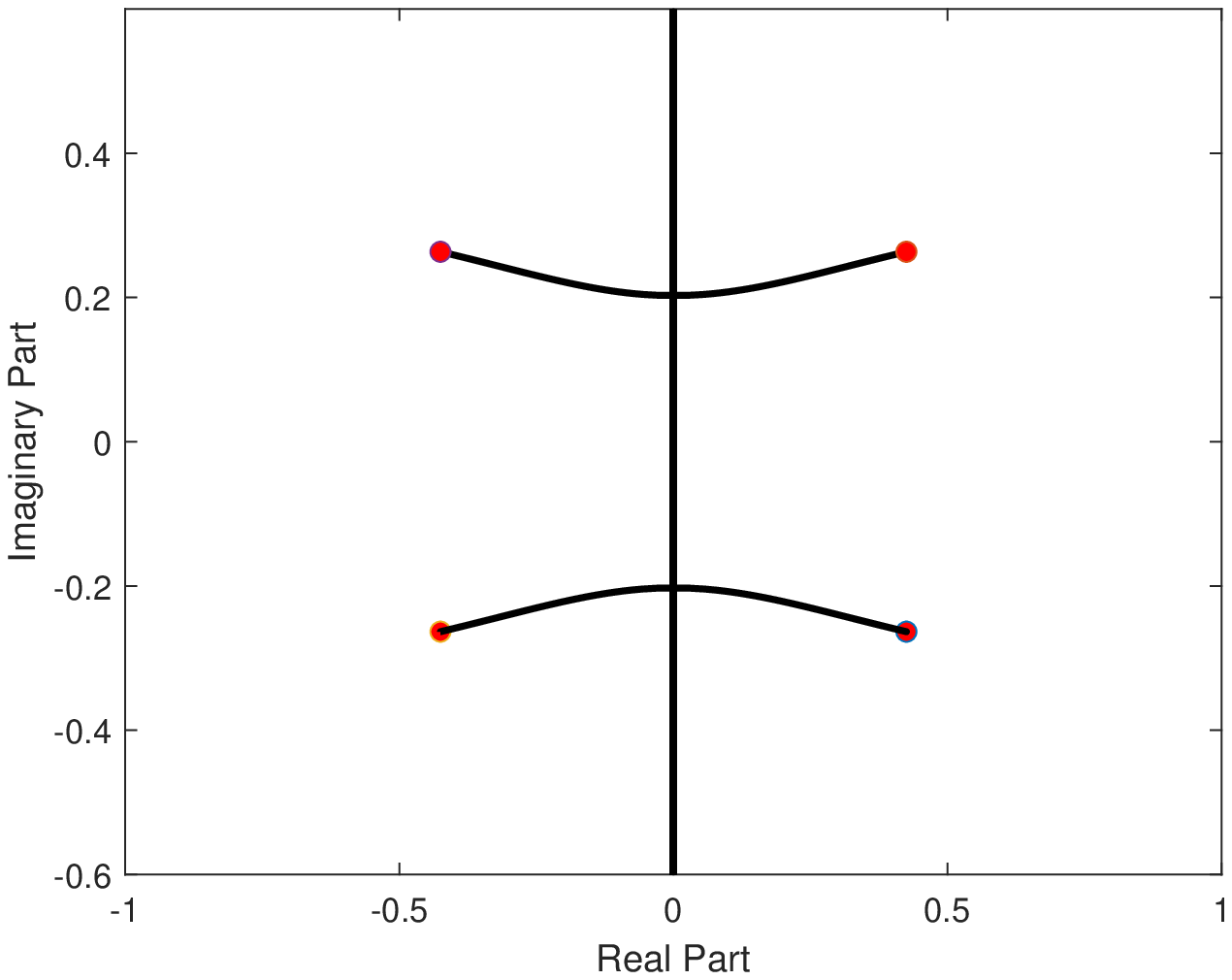}
	\includegraphics[scale=0.325]{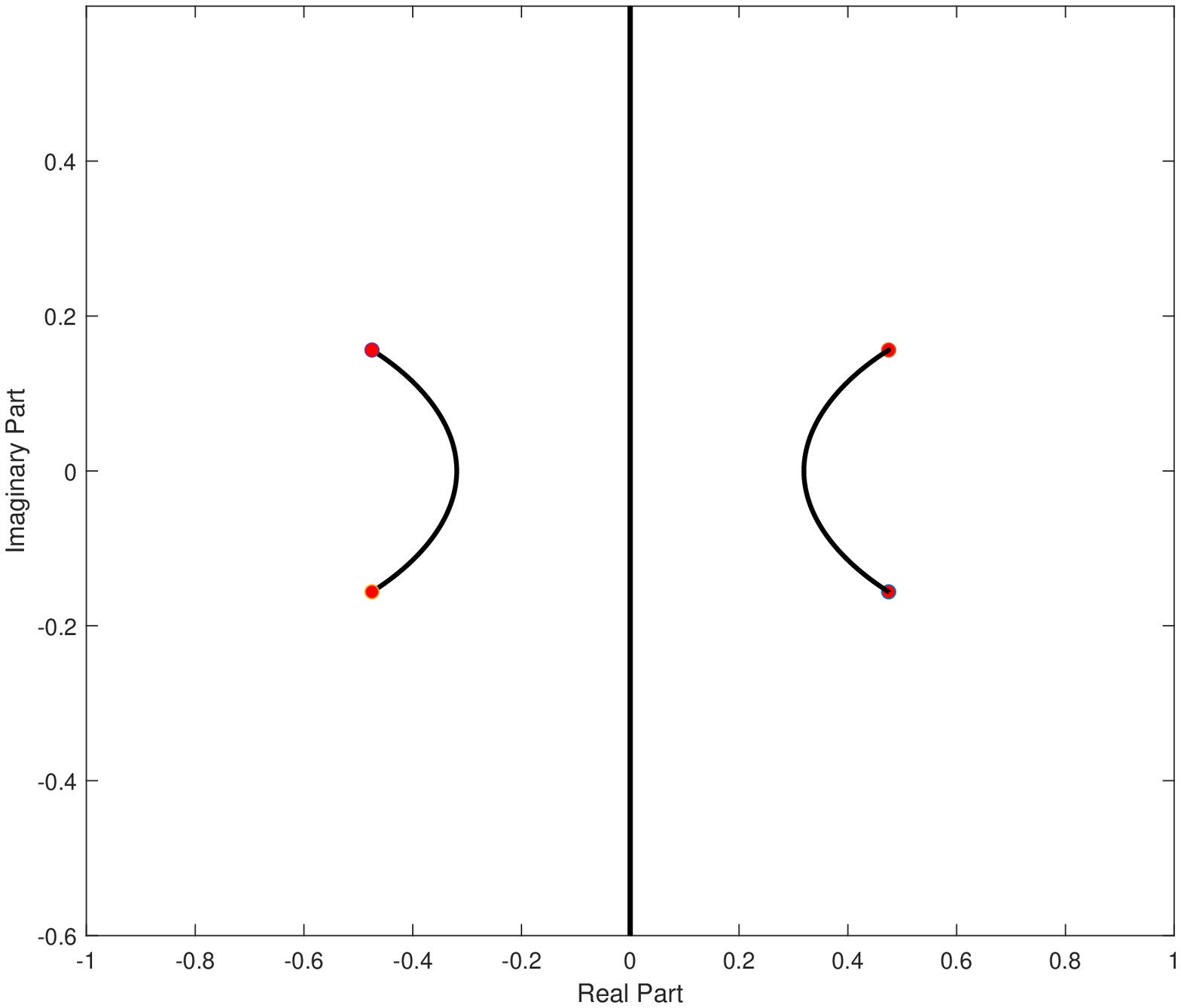}
	\caption{Lax spectrum for the ${\rm cn}$-periodic wave (\ref{red-2}) with $k=0.85$ (left) and $k=0.95$ (right). Red dots
	represent eigenvalues (\ref{eig-cn}).}
	\label{f2}
\end{figure}

Figs. \ref{f1} and \ref{f2} represent the Lax spectrum
computed numerically by using the Floquet--Bloch decomposition
of solutions to the spectral problem (\ref{3.1}) with the potentials $u$
in the form (\ref{red-1}) and (\ref{red-2}) respectively.
The black curves represent the purely continuous spectrum whereas
the red dots represent eigenvalues (\ref{eig-dn}) and (\ref{eig-cn}) respectively.
Appendix \ref{appendix-a} gives details of the Floquet--Bloch decomposition
and the numerical method used to compute the Lax spectrum at the periodic
standing waves.

\subsection{Periodic standing waves with nontrivial phase}

Let us set $c = 0$ but consider $a \neq 0$. Substituting the decomposition
$u(x) = R(x) e^{i \Theta(x)}$ with real $R$ and $\Theta$ into (\ref{Constant-2b}) and (\ref{Constant-2a})  with $c = 0$ yields
the system of first-order equations:
\begin{eqnarray}
\label{system-R-Theta}
\left\{ \begin{array}{l}
\displaystyle
\left( \frac{d R}{d x} \right)^2 + R^2 \left( \frac{d \Theta}{d x} \right)^2 + R^4 - 4b R^2 = 8d, \\
\displaystyle
R^2 \frac{d \Theta}{d x} = -2 a. \end{array} \right.
\end{eqnarray}
Substituting $\frac{d \Theta}{dx}$ from the second equation to the first equation
results in the following first-order quadrature:
\begin{equation}
\label{R-equation}
\left( \frac{dR}{dx} \right)^2 + 4 a^2 R^{-2} + R^4 - 4 b R^2 = 8d.
\end{equation}
The singularity $R = 0$ is unfolded with the transformation $\rho = R^2$ which yields
\begin{equation}
\label{Jacobi}
\frac{1}{4} \left( \frac{d \rho}{dx} \right)^2 + Z(\rho) = 0, \quad
Z(\rho) := \rho^3 - 4b \rho^2 - 8d \rho + 4 a^2.
\end{equation}
Since $\rho = R^2 \geq 0$ and $Z(0) = 4 a^2 \geq 0$, one of the roots of the cubic polynomial $Z$ is negative.
Therefore, the positive periodic solutions of the first-order quadrature (\ref{Jacobi}) exist only if there exist three real
roots of $Z$. We denote the roots by $\{\rho_1,\rho_2,\rho_3\}$ and order them as follows:
\begin{equation}
\label{root-order}
\rho_3 \leq 0 \leq \rho_2 \leq \rho_1.
\end{equation}
The connection formulas of roots $\rho_1$, $\rho_2$, and $\rho_3$ to the parameters $(b,d)$
in (\ref{quad-1}) are given by
\begin{equation}
\label{roots}
\left\{ \begin{array}{l}
4b = \rho_1 + \rho_2 + \rho_3, \\
8d = -\rho_1 \rho_2 - \rho_1 \rho_3 - \rho_2 \rho_3, \\
4a^2 = -\rho_1 \rho_2 \rho_3.
\end{array} \right.
\end{equation}
Only one square root for $a$ must be used for a particular periodic standing wave. In what follows, we shall use
the negative square root with $2a = -\sqrt{\rho_1 \rho_2} \sqrt{-\rho_3}$,
which is real-valued thanks to (\ref{root-order}).

The positive periodic solution is located in the interval $[\rho_2,\rho_1]$ and is given by
\begin{equation}
\label{Jacobi-explicit}
\rho(x) = \rho_1 - (\rho_1-\rho_2) {\rm sn}^2(\alpha x; k),
\end{equation}
where $\alpha$ and $k$ are related to $(\rho_1,\rho_2,\rho_3)$ by
\begin{equation}
\label{alpha-k}
\alpha^2 = \rho_1 - \rho_3, \quad k^2 = \frac{\rho_1 - \rho_2}{\rho_1 - \rho_3}.
\end{equation}
Thanks to the scaling transformation (\ref{scaling-transform}), we can set $\alpha = 1$ and use the
parametrization $\rho_1 = \beta$, $\rho_2 = \beta - k^2$, and $\rho_3 = \beta - 1$
which yields the exact expression considered in \cite{DS}:
\begin{equation}
\label{Jacobi-explicit-rho}
\rho(x) = \beta - k^2 {\rm sn}^2(x;k).
\end{equation}
The exact solution (\ref{Jacobi-explicit-rho}) has two parameters $\beta$ and $k$ which
belong to the following triangular region:
$k \in [0,1]$ (since $\rho_3 \leq \rho_2 \leq \rho_1$),
$\beta \leq 1$ (since $\rho_3 \leq 0$), and $\beta \geq k^2$ (since $\rho_2 \geq 0$).
On the three boundaries, we have reductions
to the ${\rm dn}$-periodic wave (\ref{red-1}) if $\beta = 1$ ($\rho_3 = 0$),
the ${\rm cn}$-periodic wave (\ref{red-2}) if $\beta = k^2$ ($\rho_2 = 0$),
and the constant-amplitude wave if $k = 0$ $(\rho_1 = \rho_2$):
\begin{eqnarray}
\label{red-3}
u(x) = \sqrt{\beta} e^{i \sqrt{1-\beta} x}, \quad \beta \in (0,1).
\end{eqnarray}
Substituting (\ref{roots}) into $P(\lambda)$ given by (\ref{Polynomial-2}) with $c = 0$ yields
\begin{equation}
\label{P-poly}
P(\lambda) = \lambda^4 - \frac{1}{2} (\rho_1 + \rho_2 + \rho_3) \lambda^2 - i \sqrt{\rho_1\rho_2} \sqrt{-\rho_3} \lambda
+ \frac{1}{16} (\rho_1^2 + \rho_2^2 + \rho_3^2 - 2 \rho_1 \rho_2 - 2 \rho_1 \rho_3 - 2 \rho_2 \rho_3).
\end{equation}
The four roots of $P(\lambda)$ can now be found in the explicit form:
\begin{equation}
\label{eig-1-2-3-4}
\lambda_1^{\pm} = \pm \frac{1}{2} (\sqrt{\rho_1} + \sqrt{\rho_2}) + \frac{i}{2} \sqrt{-\rho_3}, \quad
\lambda_2^{\pm} = \pm \frac{1}{2} (\sqrt{\rho_1} - \sqrt{\rho_2}) - \frac{i}{2} \sqrt{-\rho_3},
\end{equation}
which was stated in equation (88) in \cite{DS} without a proof.

\begin{figure}[h!]
	\centering
	\includegraphics[scale=0.5]{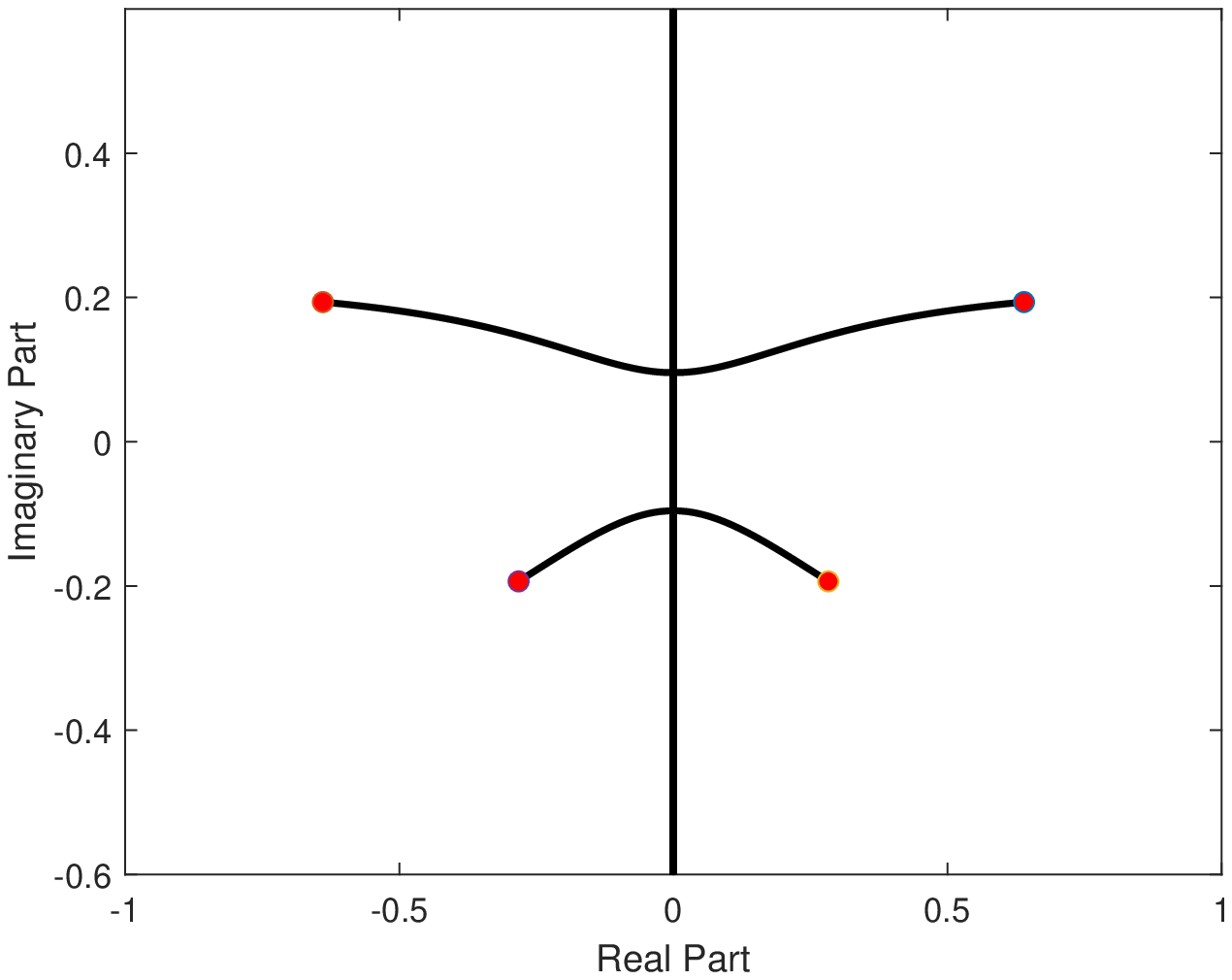}
	\includegraphics[scale=0.5]{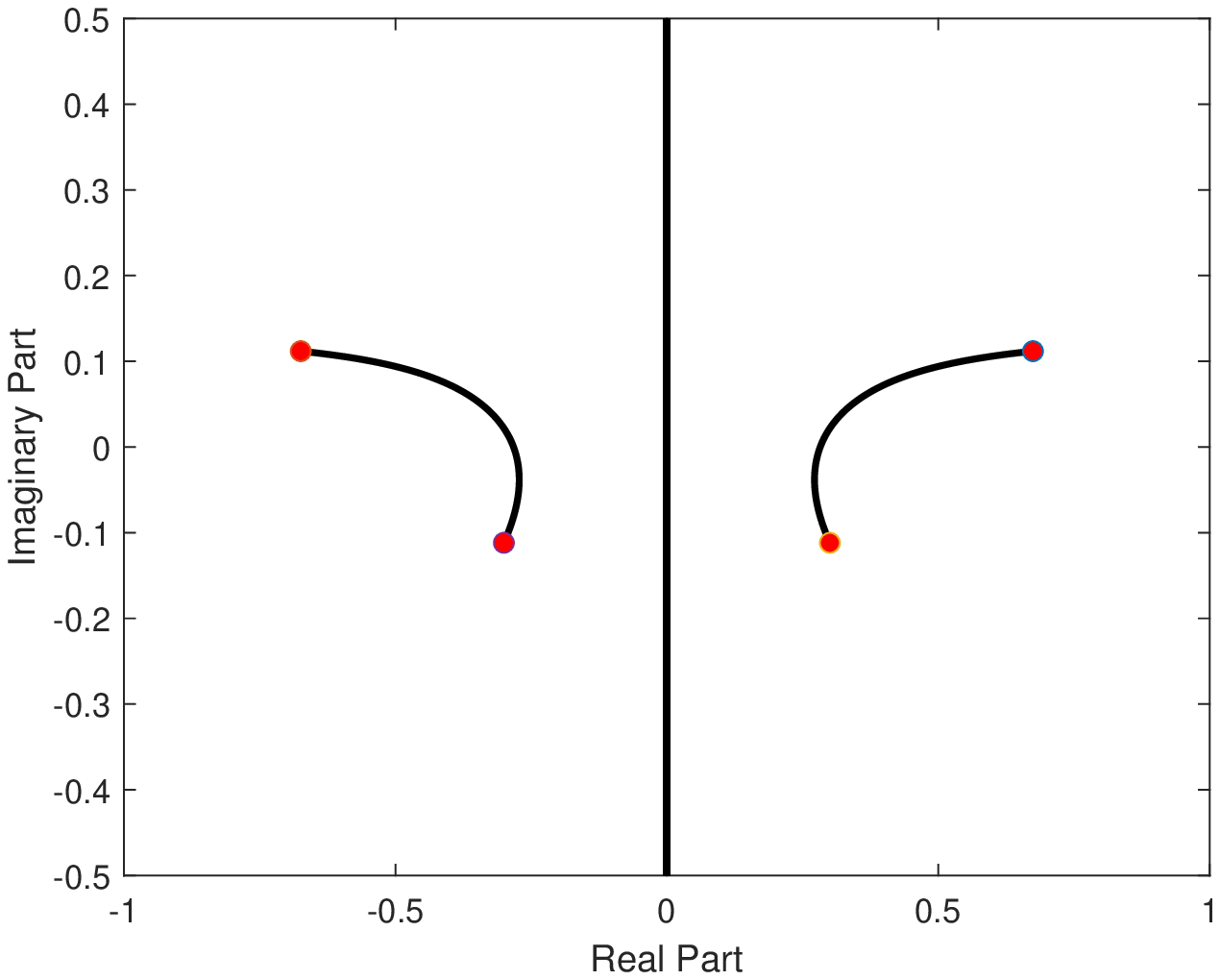}
	\caption{Lax spectrum for the periodic wave (\ref{Jacobi-explicit-rho}) with $(\beta,k) = (0.85,0.85)$ (left) and
    $(\beta,k) = (0.95,0.9)$ (right). Red dots represent eigenvalues (\ref{eig-1-2-3-4}).}
	\label{f4}
\end{figure}

Fig. \ref{f4} represents the Lax spectrum
computed numerically by using the Floquet--Bloch decomposition
of solutions to the spectral problem (\ref{3.1}) with the potentials given by $u(x) = R(x) e^{i \Theta(x)}$.
The transformation
\begin{equation}
\label{tilde-transform}
p_1(x) = \tilde{p}_1(x) e^{i \Theta(x)/2}, \quad q_1(x) = \tilde{q}_1(x) e^{-i \Theta(x)/2}
\end{equation}
is used to reduce the spectral problem (\ref{3.1}) to the one with a periodic potential considered
in Appendix \ref{appendix-a}. Note that the transformation (\ref{tilde-transform}) affects the eigenfunctions
but preserves the eigenvalues $\lambda$ in the spectral problem (\ref{3.1}).
The black curves on Fig. \ref{f4} represent the purely continuous spectrum whereas
the red dots represent eigenvalues (\ref{eig-1-2-3-4}).

\begin{figure}[h!]
	\centering
	\includegraphics[scale=0.7]{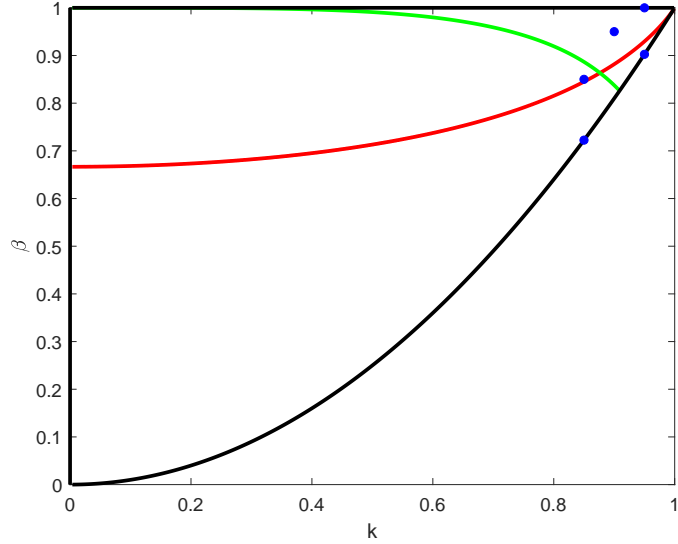}
	\caption{The black curves are boundaries of the triangular region where
the periodic waves with nontrivial phase exist.
The blue dots show parameter values of $(\beta,k)$ for the solutions
chosen for illustration on Figs. \ref{f1}, \ref{f2}, and \ref{f4}.
The green curve displays the boundary (\ref{green-boundary}), whereas
the red curve shows the boundary (\ref{boundary}).}
	\label{figure-curve}
\end{figure}

Fig. \ref{figure-curve} shows boundaries of the triangular region on the $(\beta,k)$ plane
where the periodic waves with nontrivial phase exist (black curves).
Blue dots represent the particular values of parameters $(\beta,k)$ chosen for
illustration of the Lax spectrum on Figs. \ref{f1}, \ref{f2}, and \ref{f4}.
The green curve given by the following explicit expression (see equation (100) in \cite{DS})
\begin{equation}
\label{green-boundary}
\beta = -1 + k^2 + \frac{2 E(k)}{K(k)},
\end{equation}
separates two regions on the $(\beta,k)$ plane. The Lax spectrum for the periodic waves in the lower (upper) region includes
bands crossing the imaginary (real) axis. The two choices on Figs. \ref{f2} and \ref{f4} correspond
to two points on both sides of the boundary (\ref{green-boundary}) on Fig. \ref{figure-curve}.

\section{Eigenfunctions of the linear equations}
\label{sec-eigenvectors}

Here we characterize squared eigenfunctions of the linear equations (\ref{3.1})--(\ref{3.2})
in terms of the periodic standing wave $u$. For each admissible eigenvalue $\lambda_1$ among the roots
of the polynomial $P(\lambda)$ in (\ref{Polynomial-2}), the squared eigenfunctions $p_1^2$, $\bar{q}_1^2$, and $p_1 q_1$
after the transformation (\ref{tilde-transform}) are periodic functions with the same period as the periodic wave $u$.
The second linearly independent solution of the linear equations
(\ref{3.1})--(\ref{3.2}) exists for the same eigenvalues
and is characterized in terms of the periodic eigenfunctions. The second
solution is not periodic and grows linearly in $x$ and $t$ almost everywhere on the $(x,t)$-plane.

Let us recall the representations (\ref{W-11-12}) and (\ref{3.19}) for $W_{11}(\lambda)$ and $W_{12}(\lambda)$
in terms of the squared eigenfunctions for (\ref{W-11-12}) and the periodic standing wave $u$ for (\ref{3.19}).
By evaluating the residue terms at simple poles $\lambda = \lambda_1$ and $\lambda = -\bar{\lambda}_1$
in each representation (\ref{W-11-12}) and (\ref{3.19}),
we obtain the following explicit expressions for $p_1^2$, $\bar{q}_1^2$, and $p_1 q_1$:
\begin{equation}
\label{squared-1}
\left\{ \begin{array}{l}
p_1^2 = \frac{1}{\lambda_1 + \bar{\lambda}_1} \left[ \frac{1}{2} \frac{du}{dx} + i c u + \lambda_1 u \right], \\
\bar{q}_1^2 = \frac{1}{\lambda_1 + \bar{\lambda}_1} \left[ -\frac{1}{2} \frac{du}{dx} - i c u + \bar{\lambda}_1 u \right], \\
p_1 q_1 = -\frac{1}{\lambda_1 + \bar{\lambda}_1} \left[ \frac{1}{2} |u|^2 - b + i \lambda_1 c + \lambda_1^2 \right]. \end{array} \right.
\end{equation}

Let $\varphi = (p_1,q_1)^T$ be a solution of the linear equations (\ref{3.1})--(\ref{3.2}) for $\lambda = \lambda_1$.
If $u$ is a $L$-periodic standing wave, then $p_1^2$ and $q_1^2$ is periodic with the same period $L$ in $x$ after the transformation (\ref{tilde-transform}).
The second, linearly independent solution $\varphi = (\hat{p}_1,\hat{q}_1)^T$ of the same equations
(\ref{3.1})--(\ref{3.2}) for $\lambda = \lambda_1$ can be written in the form:
\begin{equation}
\label{represent-new}
\hat{p}_1 = p_1 \phi_1 - \frac{2 \bar{q}_1}{|p_1|^2 + |q_1|^2}, \quad
\hat{q}_1 = q_1 \phi_1 + \frac{2 \bar{p}_1}{|p_1|^2 + |q_1|^2},
\end{equation}
where $\phi_1$ is to be determined as a function of $(x,t)$. Wronskian between the two solutions is normalized by
$p_1 \hat{q}_1 - \hat{p}_1 q_1 = 2$. Note that if $u$ is the periodic standing wave in the form
\begin{equation}
u(x,t) = U(x) e^{2ibt},
\label{separation-U}
\end{equation}
then the first solution $\varphi = (p_1,q_1)^T$ to the linear equations (\ref{3.1})--(\ref{3.2}) for $\lambda = \lambda_1$
is written in the form
\begin{equation}
\label{separation}
p_1(x,t) = P_1(x) e^{ibt}, \quad q_1(x,t) = Q_1(x) e^{-ibt},
\end{equation}
whereas the second solution $\varphi = (\hat{p}_1,\hat{q}_1)^T$ to the same equations
is written in the form
\begin{equation}
\label{separation-hat}
\hat{p}_1(x,t) = \hat{P}_1(x,t) e^{ibt}, \quad \hat{q}_1(x,t) = \hat{Q}_1(x,t) e^{-ibt},
\end{equation}
where $\hat{P}_1(x,t)$ and $\hat{Q}_1(x,t)$ depend on $t$ only through the function $\phi_1(x,t)$.

Substituting (\ref{represent-new}) into (\ref{3.1})
and using (\ref{3.1}) for $\varphi = (p_1,q_1)^T$ yields the following equation
for $\phi_1$:
\begin{equation}
\label{phi-der-x}
\frac{\partial \phi_1}{\partial x} = -\frac{4 (\lambda_1 + \bar{\lambda}_1) \bar{p}_1 \bar{q}_1}{(|p_1|^2+|q_1|^2)^2}.
\end{equation}
Similarly, substituting (\ref{represent-new}) into (\ref{3.2})
and using (\ref{3.2}) for $\varphi = (p_1,q_1)^T$ yields another equation for
$\phi_1$:
\begin{equation}
\label{phi-der-t}
\frac{\partial \phi_1}{\partial t} = -\frac{4 i (\lambda_1^2 - \bar{\lambda}_1^2) \bar{p}_1 \bar{q}_1}{(|p_1|^2+|q_1|^2)^2}
+ \frac{2i (\lambda_1 + \bar{\lambda}_1) (u \bar{p}_1^2 + \bar{u} \bar{q}_1^2)}{(|p_1|^2+|q_1|^2)^2}.
\end{equation}
The system of first-order equations (\ref{phi-der-x}) and (\ref{phi-der-t}) is compatible in the sense
$\phi_{1xt} = \phi_{1tx}$ since it is derived from the compatible Lax system (\ref{3.1})--(\ref{3.2}).
Note also that the transformation (\ref{separation}) leaves equations (\ref{phi-der-x}) and (\ref{phi-der-t})
invariant for $\phi_1(x,t)$.

\subsection{Periodic standing waves with trivial phase}

We set $c = a = 0$ in solutions of the system (\ref{LN-2}), (\ref{Constant-2b}), and (\ref{Constant-2a}).
We also use the scaling transformation (\ref{scaling-transform}). There exist
two admissible pairs of values of $\lambda_1$ given by (\ref{eig-dn}) for the ${\rm dn}$-periodic solution (\ref{red-1})
and (\ref{eig-cn}) for the ${\rm cn}$-periodic solution (\ref{red-2}).
Let us fix
\begin{equation}
\label{lambda-1-dn-cn}
\lambda_1 = \frac{1}{2} (u_1 \pm u_2),
\end{equation}
for two eigenvalues in either (\ref{eig-dn}) or (\ref{eig-cn}).

The periodic standing waves are written in the form (\ref{separation-U}) with real-valued $U$.
By separating the variables in (\ref{separation}) and using $b = \frac{1}{4} (u_1^2 + u_2^2)$, we obtain
\begin{eqnarray}
\left\{ \begin{array}{l}
P_1^2 = \frac{1}{\lambda_1 + \bar{\lambda}_1} \left( \frac{1}{2} \frac{dU}{dx} + \lambda_1 U \right), \\
\bar{Q}_1^2 = \frac{1}{\lambda_1 + \bar{\lambda}_1} \left( -\frac{1}{2} \frac{dU}{dx} + \bar{\lambda}_1 U \right), \\
P_1 Q_1 = -\frac{1}{2(\lambda_1 + \bar{\lambda}_1)} \left( \pm u_1 u_2 + U^2 \right).\end{array} \right.
\label{pq-2}
\end{eqnarray}

The quantities $\lambda_1$, $P_1$ and $Q_1$ are real-valued for the ${\rm
dn}$-periodic wave (\ref{red-1}). By using (\ref{pq-2}), we solve the
first-order equations (\ref{phi-der-x}) and (\ref{phi-der-t}) in the closed
form:
\begin{equation}
\label{phi-2}
\phi_1(x,t) = 2x + 2 i (1 \pm \sqrt{1-k^2}) t \pm 2 \sqrt{1-k^2} \int_0^x \frac{dy}{{\rm dn}^2(y;k)}
\end{equation}
up to addition of a complex-valued constant. We note that
\begin{equation}
\label{inequality-K-E}
K(k) \pm \sqrt{1-k^2} \int_0^{K(k)} \frac{dy}{{\rm dn}^2(y;k)} = K(k) \pm \frac{E(k)}{\sqrt{1-k^2}} \gtrless 0, \quad k \in (0,1),
\end{equation}
where the inequality for the upper sign is trivial, whereas the inequality for the lower sign follows
from the inequality 19.9.8 in \cite{Olver}, that is, from
\begin{equation}
\label{E-K-inequality}
\frac{E(k)}{K(k)} > \sqrt{1 - k^2}.
\end{equation}
Thanks to (\ref{inequality-K-E}), the function $\phi_1(x,t)$ grows linearly as $|x| + |t| \to \infty$ for both signs.
Hence, the second solution $\varphi = (\hat{p}_1,\hat{q}_1)^T$ given by (\ref{represent-new})
with $\phi_1$ in (\ref{phi-2}) grows linearly as $|x| + |t| \to \infty$ everywhere on the $(x,t)$-plane.
Compared to the representation for $\varphi = (\hat{p}_1,\hat{q}_1)^T$
used in \cite{CPnls}, the new representation (\ref{represent-new})
has the advantage of being equally applicable to both eigenvalues $\lambda_1 = \frac{1}{2} (u_1 \pm u_2)$
because the denominators in the new representation (\ref{represent-new})
with (\ref{phi-2}) never vanish.

For the ${\rm cn}$-periodic wave (\ref{red-2}), $\lambda_1$ is complex and so are $P_1$ and $Q_1$.
It was found in \cite{CPnls} that
\begin{equation}
\label{cn-identity-1}
|P_1(x)|^2 + |Q_1(x)|^2 = {\rm dn}(x;k)
\end{equation}
and
\begin{equation}
\label{cn-identity-2}
2 P_1(x) \bar{Q}_1(x) =  - {\rm cn}(x;k) {\rm dn}(x;k) \mp i \sqrt{1-k^2} {\rm sn}(x;k).
\end{equation}
By using (\ref{pq-2}) and (\ref{cn-identity-1}), we solve the first-order equations (\ref{phi-der-x}) and (\ref{phi-der-t})
in the closed form:
\begin{eqnarray}
\phi_1(x,t) = 2 k^2 \int_0^{x} \frac{{\rm cn}^2(y;k) dy}{{\rm dn}^2(y;k)} \mp
2 i k \sqrt{1-k^2} \int_0^{x} \frac{dy}{{\rm dn}^2(y;k)} + 2 i k t
\label{phi-2-cn}
\end{eqnarray}
up to addition of a complex-valued constant. It is clear from (\ref{phi-2-cn})
that $\phi_1(x,t)$ grows linearly as $|x| + |t| \to \infty$ for both signs in (\ref{phi-2-cn}), hence,
the second solution $\varphi = (\hat{p}_1,\hat{q}_1)^T$ given by (\ref{represent-new}) with $\phi_1$ in
(\ref{phi-2-cn}) also grows linearly as $|x| + |t| \to \infty$ everywhere on the $(x,t)$-plane.
Compared to the representation used in \cite{CPnls}, it is now easier to see the growth of
$\phi_1(x,t)$ at infinity.

\subsection{Periodic standing waves with nontrivial phase}

We set $c = 0$ for solutions of the system (\ref{LN-2}), (\ref{Constant-2b}), and (\ref{Constant-2a}).
Let the roots $\{ \rho_1,\rho_2,\rho_3\}$ satisfy the order (\ref{root-order}).
There exist two admissible pairs of values of $\lambda_1$ and they
are given in the explicit form (\ref{eig-1-2-3-4}). Let us fix
\begin{equation}
\label{lambda-3}
\lambda_1 = \frac{1}{2} (\sqrt{\rho_1} \pm \sqrt{\rho_2}) \pm \frac{i}{2} \sqrt{-\rho_3}
\end{equation}
for two eigenvalues in (\ref{eig-1-2-3-4}).

The periodic standing waves are given in the form (\ref{separation-U}) with complex-valued $U$.
By separating the variables in (\ref{separation}) and
using $b = \frac{1}{4} (\rho_1 + \rho_2 + \rho_3)$, we obtain
\begin{eqnarray}
\left\{ \begin{array}{l}
P_1^2 = \frac{1}{\lambda_1 + \bar{\lambda}_1} \left( \frac{1}{2} \frac{dU}{dx} + \lambda_1 U \right), \\
\bar{Q}_1^2 = \frac{1}{\lambda_1 + \bar{\lambda}_1} \left( -\frac{1}{2} \frac{dU}{dx} + \bar{\lambda}_1 U \right), \\
P_1 Q_1 = -\frac{1}{2(\lambda_1 + \bar{\lambda}_1)} \left( \pm \sqrt{\rho_1 \rho_2} \pm i
\sqrt{-\rho_3} (\sqrt{\rho_1} \pm \sqrt{\rho_2}) + |U|^2 \right).\end{array} \right.
\label{pq-3}
\end{eqnarray}
Next, we represent the solutions in the polar form:
\begin{equation}
\label{separation-3-new}
U(x) = R(x) e^{i \Theta(x)}, \quad P_1(x) = \tilde{P}_1(x) e^{i \Theta(x)/2}, \quad Q_1(x) = \tilde{Q}_1(x) e^{-i\Theta(x)/2},
\end{equation}
where $R$ and $\Theta$ are real, whereas $\tilde{P}_1$ and $\tilde{Q}_1$ are complex-valued.
In Appendix \ref{appendix-b}, we prove that
\begin{equation}
\label{dn-relation-1}
|\tilde{P}_1(x)|^2 + |\tilde{Q}_1(x)|^2 = {\rm dn}(x;k)
\end{equation}
and
\begin{equation}
\label{dn-relation-2}
\tilde{P}_1(x) \overline{\tilde{Q}}_1(x) = -\frac{1}{2(\sqrt{\rho}_1 \pm \sqrt{\rho}_2)}
\left[ \frac{{\rm dn}(x;k)}{R(x)} \left( R(x)^2 \pm \sqrt{\rho_1 \rho_2} \right)
\mp \frac{i \sqrt{-\rho_3}}{{\rm dn}(x;k)} \frac{dR}{dx}\right].
\end{equation}
By using (\ref{pq-3}) and (\ref{dn-relation-1}), we solve the first-order equations (\ref{phi-der-x}) and (\ref{phi-der-t})
in the closed form:
\begin{equation}
\label{phi-3}
\phi_1(x,t) = 2 \int_0^x \frac{\rho(y) \pm \sqrt{\rho_1 \rho_2} \mp i \sqrt{-\rho_3} (\sqrt{\rho_1}
\pm \sqrt{\rho}_2)}{{\rm dn}^2(y;k)} dy + 2 i (\sqrt{\rho_1} \pm \sqrt{\rho_2}) t
\end{equation}
up to addition of a complex-valued constant. When $\rho_1 = 1$, $\rho_2 = 1 - k^2$, and $\rho_3 = 0$,
expression (\ref{phi-3}) is equivalent to (\ref{phi-2}) for the ${\rm dn}$-periodic wave (\ref{red-1}).
When $\rho_1 = k^2$, $\rho_2 = 0$, and $\rho_3 = -(1-k^2)$,
expression (\ref{phi-3}) is equivalent to (\ref{phi-2-cn}) for the ${\rm cn}$-periodic wave (\ref{red-2}).

It is clear from (\ref{phi-3}) with the upper sign that $\phi_1(x,t)$ grows linearly as $|x| + |t| \to \infty$.
On the other hand, it follows from (\ref{phi-3}) with the lower sign that $\phi_1(x,t)$
does not grow everywhere as $|x| + |t| \to \infty$ if
\begin{equation}
\label{boundary-preliminary}
\int_0^{K(k)} \frac{\rho(y) - \sqrt{\rho_1 \rho_2}}{{\rm dn}^2(y;k)} dy = 0,
\end{equation}
where the integral for $\rho$ given by (\ref{Jacobi-explicit-rho})
can be evaluated in Jacobian elliptic integrals as follows:
\begin{equation}
\label{boundary}
K(k) + \frac{\beta - 1 - \sqrt{\beta (\beta - k^2)}}{1 - k^2} E(k) = 0.
\end{equation}
If the constraint (\ref{boundary-preliminary}) is satisfied, then $\phi_1(x,t)$ is bounded as $|x| + |t| \to \infty$
along the family of straight lines
\begin{equation}
\label{straight-line}
t + \sqrt{-\rho_3} \left[ \int_0^{K(k)} \frac{dy}{{\rm dn}^2(y;k)} \right] \frac{x}{K(k)} = t_0,
\end{equation}
where $t_0 \in \mathbb{R}$ is arbitrary.

We claim that there exists exactly one root of the constraint (\ref{boundary}) in $\beta \in (k^2,1)$ for every $k \in (0,1)$.
Indeed, we obtain
\begin{eqnarray*}
\beta = k^2 : & \quad & K(k) + \frac{\beta - 1 - \sqrt{\beta (\beta - k^2)}}{1 - k^2} \biggr|_{\beta = k^2} E(k) = K(k) - E(k) > 0, \\
\beta = 1 : & \quad & K(k) + \frac{\beta - 1 - \sqrt{\beta (\beta - k^2)}}{1 - k^2} \biggr|_{\beta = 1} E(k) =
K(k) - \frac{E(k)}{\sqrt{1 - k^2}} < 0,
\end{eqnarray*}
where the first inequality is due to 19.9.6 in \cite{Olver} and
the second inequality is due to (\ref{E-K-inequality}). Since the left-hand side
of (\ref{boundary}) is a $C^1$ function of $\beta$ in $(k^2,1)$ for every $k \in (0,1)$,
there exists at least one $\beta \in (k^2,1)$ such that constraint (\ref{boundary}) is satisfied.
Moreover, differentiating the left-hand side of (\ref{boundary}) in $\beta$ yields for $\beta \in [k^2,1]$:
\begin{eqnarray*}
\frac{E(k)}{1-k^2} \left[ 1 - \frac{\sqrt{\beta - k^2}}{2 \sqrt{\beta}} - \frac{\sqrt{\beta}}{2 \sqrt{\beta - k^2}} \right]
= \frac{E(k)}{2(1-k^2) \sqrt{\beta (\beta - k^2)}} \left[ 2 \sqrt{\beta(\beta - k^2)} - \beta - (\beta - k^2) \right],
\end{eqnarray*}
where the right-hand side is negative due to the Cauchy--Schwarz inequality.
Hence, the left-hand side of (\ref{boundary}) is decreasing in $\beta$ and
there is exactly one $\beta \in (k^2,1)$ where the constraint (\ref{boundary}) is satisfied.
The red curve on Fig. \ref{figure-curve} shows the only root of the constraint (\ref{boundary}) on the $(\beta,k)$ plane.

\section{Rogue waves on the periodic background}

Here we use the one-fold Darboux transformation (\ref{1-fold}) with the second solution
$\varphi = (\hat{p}_1,\hat{q}_1)^t$ of the linear equations (\ref{3.1})--(\ref{3.2}) with $\lambda = \lambda_1$:
\begin{equation}
\label{1-fold-rogue}
\hat{u} = u + \frac{2 (\lambda_1 + \bar{\lambda}_1) \hat{p}_1 \bar{\hat{q}}_1}{|\hat{p}_1|^2 + |\hat{q}_1|^2},
\end{equation}
where $u$ is the periodic standing wave in the form (\ref{separation-U})
and $\varphi = (\hat{p}_1,\hat{q}_1)^T$ is a solution to the linear equations
(\ref{3.1})--(\ref{3.2}) for $\lambda = \lambda_1$ written in the form (\ref{represent-new}).
Substituting (\ref{represent-new}) into (\ref{1-fold-rogue}) yields a more explicit formula:
\begin{equation}
\label{1-fold-rogue-explicit}
\hat{u} = \tilde{u} + \frac{4 (\lambda_1 + \bar{\lambda}_1) \left[ (p_1^2 \phi_1 - \bar{q}_1^2 \bar{\phi}_1) (|p_1|^2 + |q_1|^2) -
4 p_1 \bar{q}_1\right]}{(|p_1|^2 + |q_1|^2) \left[ 4 + |\phi_1|^2 (|p_1|^2 + |q_1|^2)^2 \right]},
\end{equation}
where $\varphi = (p_1,q_1)^T$ is a periodic solution to the linear equations
(\ref{3.1})--(\ref{3.2}) for $\lambda = \lambda_1$ in the form (\ref{separation}),
$\phi_1$ is a linearly growing solution to the system (\ref{phi-der-x}) and (\ref{phi-der-t}),
and $\tilde{u}$ is given by
\begin{equation}
\label{1-fold-rogue-limit}
\tilde{u} = u + \frac{2 (\lambda_1 + \bar{\lambda}_1) p_1
\bar{q}_1}{|p_1|^2 + |q_1|^2}.
\end{equation}
We will show with explicit analytical computations that $\tilde{u}$ is a translated version of $u$
along symmetries of the NLS equation (\ref{nls}). It also follows from
(\ref{phi-2}), (\ref{phi-2-cn}), and (\ref{phi-3}) that $|\phi_1(x,t)| \to \infty$
as $|x| + |t| \to \infty$ everywhere on the $(x,t)$-plane with the only exception
for the periodic standing wave with non-trivial phase, parameters of which satisfy the constraint (\ref{boundary}).
Since the representation (\ref{1-fold-rogue-explicit}) implies that
$\hat{u} |_{|\phi_1| \to \infty} = \tilde{u}$, the aforementioned properties
imply that the one-fold transformation (\ref{1-fold-rogue}) generates a rogue wave
$\hat{u}$ on the background of the periodic standing wave $u$
and the rogue wave satisfies the limits (\ref{rogue-wave-def})
with the only exception given by the constraint (\ref{boundary}).

The representation (\ref{1-fold-rogue-explicit}) also implies that
\begin{equation}
\label{1-fold-rogue-origin}
\hat{u} |_{\phi_1 = 0} = u - \frac{2 (\lambda_1 + \bar{\lambda}_1) p_1 \bar{q}_1}{|p_1|^2 + |q_1|^2}
= 2 u - \tilde{u}.
\end{equation}
We will show numerically that $|\hat{u}|$ has a global maximum at $(x,t) = (0,0)$,
from which we can compute the magnification factor $M$ of the rogue wave $\hat{u}$
according to the definition (\ref{factor-magnification}). Since $\phi_1(0,0) = 0$ by the choice
of the integration constant in (\ref{phi-2}), (\ref{phi-2-cn}), and (\ref{phi-3}),
the magnification factor $M$ is computed from (\ref{1-fold-rogue-origin}) as follows:
\begin{equation}
\label{factor-magnification-explicit}
M = \frac{|\hat{u}(0,0)|}{\max_{(x,t) \in \mathbb{R}^2} |\hat{u}| |_{|\phi_1| \to \infty}} =
\frac{|2u(0,0) - \tilde{u}(0,0)|}{\max_{(x,t) \in \mathbb{R}^2} |\tilde{u}(x,t)|} \leq 3,
\end{equation}
where in the last bound we have used the property that $\tilde{u}$ is a translated version of $u$.
Hence, the magnification factor $M$ of the rogue waves obtained with the one-fold Darboux transformation
(\ref{1-fold-rogue}) does not exceed the triple magnification factor of the canonical rogue wave (\ref{rogue-basic})
attained at the constant-amplitude wave background.

\subsection{Periodic standing waves with trivial phase}

\begin{figure}[h!]
	\centering
	\includegraphics[scale=0.5]{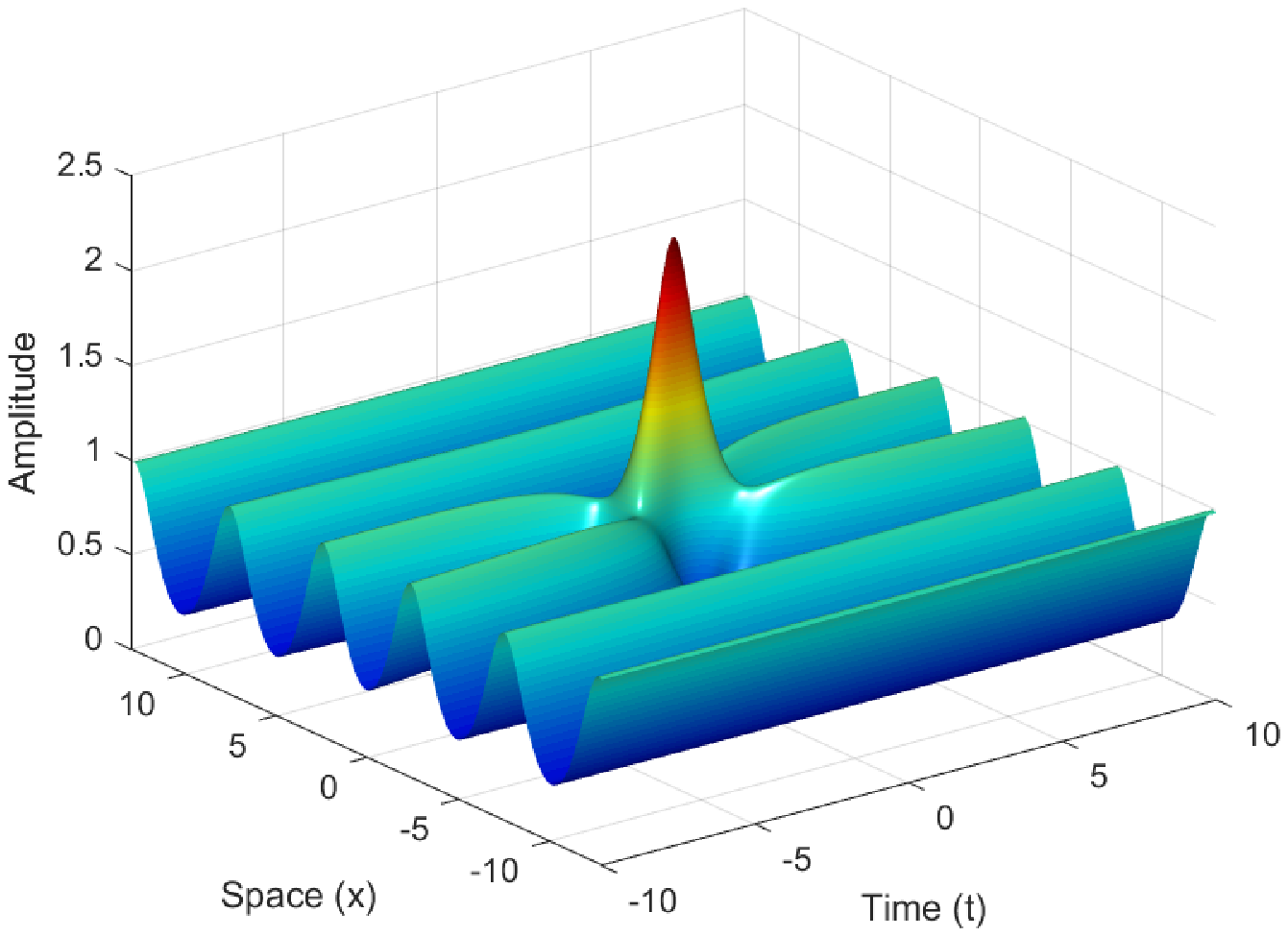}
	\includegraphics[scale=0.5]{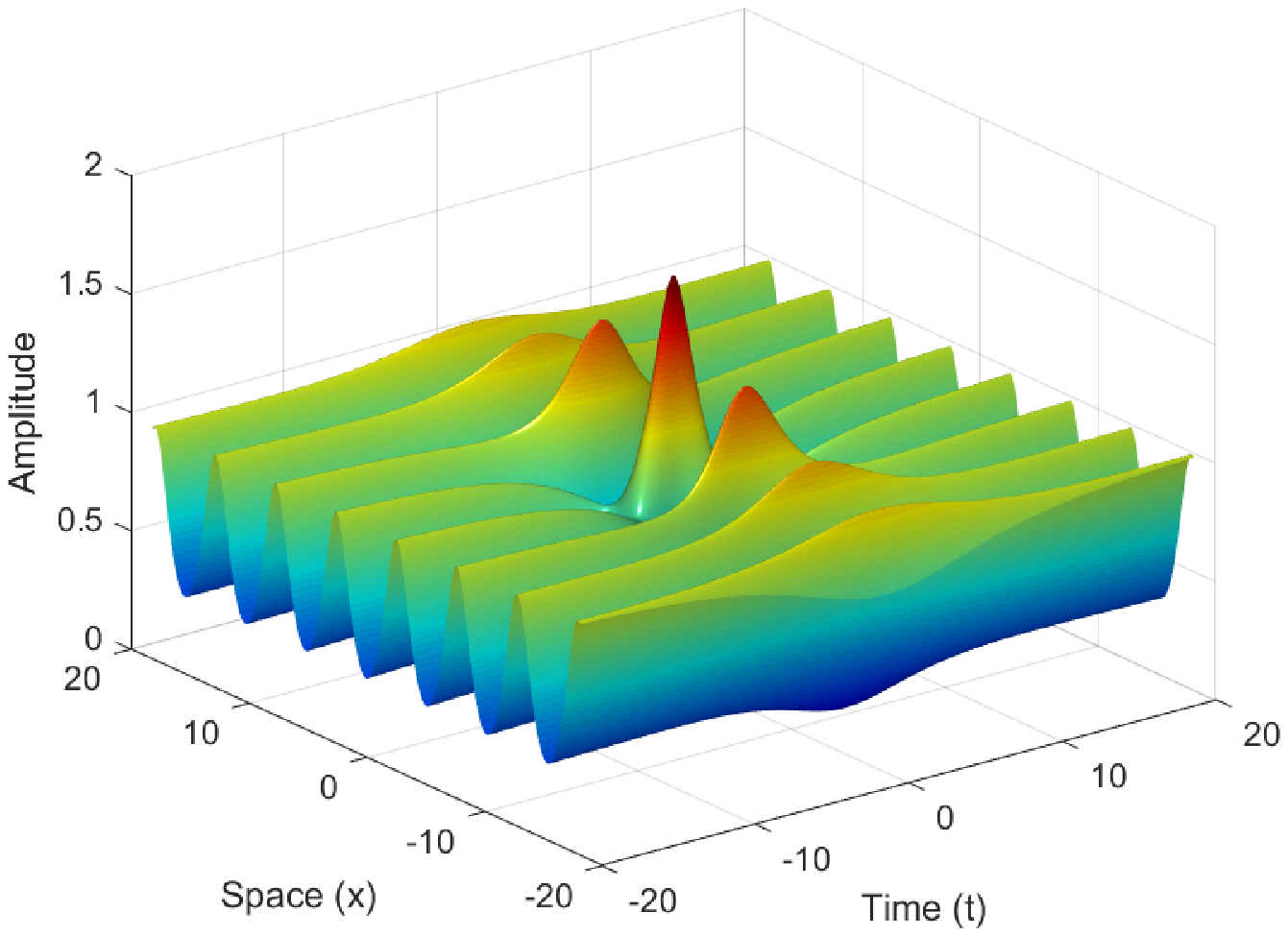}
	\caption{Rogue waves on the ${\rm dn}$-periodic wave with $k = 0.95$
    for eigenvalues $\lambda_1 = \frac{1}{2} (u_1 \pm u_2)$ with upper sign (left) and lower sign (right).}
	\label{f9}
\end{figure}

For the ${\rm dn}$-periodic wave (\ref{red-1}), we obtain from (\ref{separation-U}), (\ref{separation}), (\ref{pq-2}),
and (\ref{1-fold-rogue-explicit}):
\begin{equation}
\label{rogue-2}
\hat{u} = \left[ \mp \tilde{U} + \frac{2 (\phi_1 + \bar{\phi}_1) \frac{dU}{dx}
+ 4 \lambda_1 (\phi_1 - \bar{\phi}_1) U + 8 (U \pm \tilde{U})}{4 + |\phi_1|^2 U^2} \right] e^{2 i b t},
\end{equation}
where $U(x) = {\rm dn}(x;k)$,
$$
\tilde{U}(x) = \frac{\sqrt{1-k^2}}{{\rm dn}(x;k)} = U(x + K(k)),
$$
as follows from Table 22.4.3 in \cite{Olver}, and $\phi_1(x,t)$ is given by (\ref{phi-2}).
Fig. \ref{f9} shows rogue waves (\ref{rogue-2}) for both signs which correspond
to two choices of $\lambda = \frac{1}{2} (u_1 \pm u_2)$ with $u_1 = 1$ and $u_2 = \sqrt{1-k^2}$.

Since $|\phi_1(x,t)| \to \infty$ as $|x| + |t| \to \infty$ everywhere on the $(x,t)$ plane, we have
$\hat{u}(x,t) \sim \mp \tilde{U}(x) e^{2ibt}$ as $|x| + |t| \to \infty$, which is a translation of
the original periodic standing wave $u(x,t) = U(x) e^{2 i b t}$ in $x$ by a half-period.
On the other hand, since $\phi_1(0,0) = 0$ and the maximum of $|\hat{u}(x,t)|$ occurs at the origin
$(x,t) = (0,0)$ (see Fig. \ref{f9}), it follows from (\ref{factor-magnification-explicit}) that
\begin{equation}
\label{factor-dn}
M = \frac{|2U(0) - \tilde{U}(0)|}{\max_{x \in \mathbb{R}} |\tilde{U}(x)|} = 2 \pm \sqrt{1-k^2},
\end{equation}
which is the magnification factor of the rogue wave on the ${\rm dn}$-periodic wave (\ref{red-1})
obtained in \cite{CPnls}.

\begin{figure}[h!]
	\centering
	\includegraphics[scale=0.5]{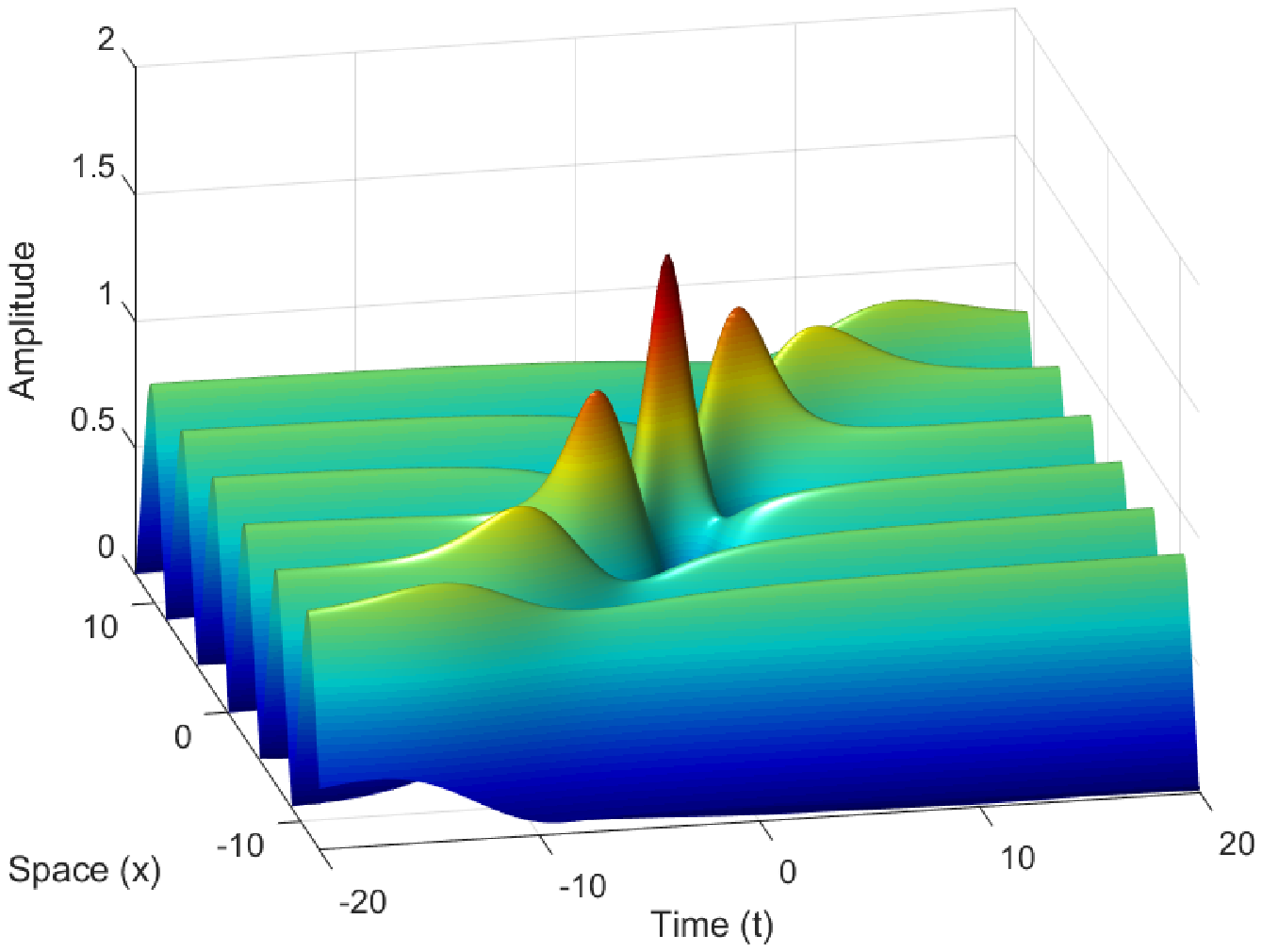}
	\includegraphics[scale=0.5]{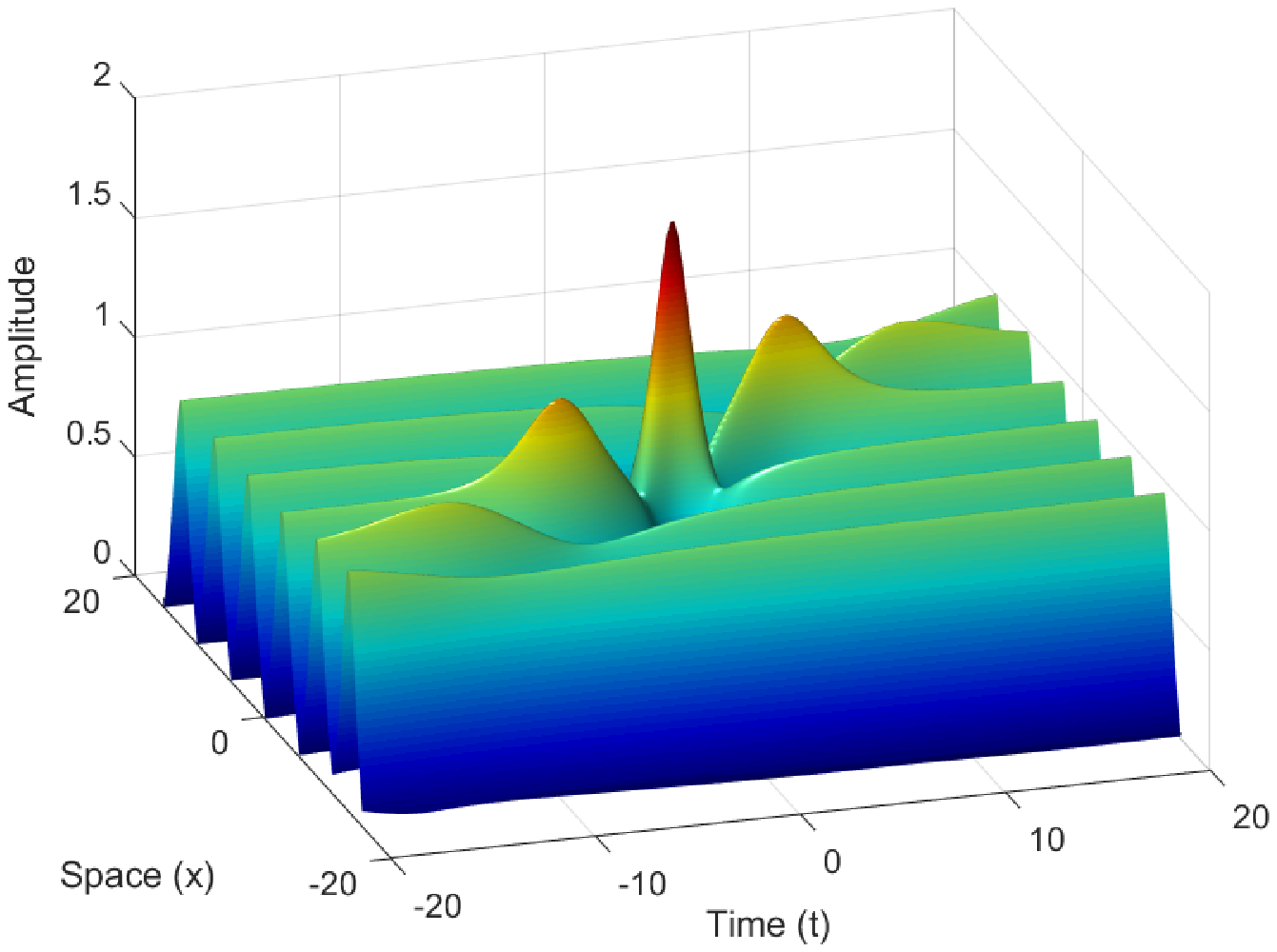}
	\caption{Rogue waves on the ${\rm cn}$-periodic wave with $k = 0.85$ (left) and $k = 0.95$ (right).}
	\label{f10}
\end{figure}

For the ${\rm cn}$-periodic wave (\ref{red-2}), we obtain from (\ref{separation-U}),
(\ref{separation}), (\ref{pq-2}), (\ref{cn-identity-1}),
(\ref{cn-identity-2}), and (\ref{1-fold-rogue-explicit}):
\begin{equation}
\label{rogue-2-cn}
\hat{u} = \left[ \pm i \tilde{U} + \frac{2 (\phi_1 + \bar{\phi}_1) \frac{dU}{dx} +
4 (\lambda_1 \phi_1 - \bar{\lambda}_1 \bar{\phi}_1) U + 8 (U \mp i \tilde{U})}{4 + |\phi_1|^2 {\rm dn}^2(x;k)} \right] e^{2 i b t},
\end{equation}
where $U(x) = k {\rm cn}(x;k)$,
$$
\tilde{U}(x) = -\frac{k \sqrt{1-k^2} {\rm sn}(x;k)}{{\rm dn}(x;k)} = U(x + K(k)),
$$
as follows from Table 22.4.3 in \cite{Olver}, and $\phi_1(x,t)$ is given by (\ref{phi-2-cn}).
Fig. \ref{f10} shows rogue wave (\ref{rogue-2-cn}) with the upper sign for two choices of $k$
with qualitatively different Lax spectrum on Fig. \ref{f2}. The two signs in (\ref{rogue-2-cn})
correspond to two choices of $\lambda = \frac{1}{2} (u_1 \pm i \nu_2)$
with $u_1 = k$ and $\nu_2 = \sqrt{1-k^2}$. The rogue wave (\ref{rogue-2-cn}) with the lower sign
propagates to the opposite direction on the $(x,t)$ plane compared to the rogue wave with the upper sign
on Fig. \ref{f10}.

Since $|\phi_1(x,t)| \to \infty$ as $|x| + |t| \to \infty$
everywhere on the $(x,t)$ plane, we have
$\hat{u}(x,t) \sim \pm i \tilde{U}(x) e^{2ibt}$ as $|x| + |t| \to \infty$, which is a translation of
the original periodic standing wave $u(x,t) = U(x) e^{2 i b t}$ in $x$ by
a quarter period. On the other hand, since $\phi_1(0,0) = 0$ and the maximum of $|\hat{u}(x,t)|$ occurs at the origin
$(x,t) = (0,0)$ (see Fig. \ref{f10}), it follows from (\ref{factor-magnification-explicit}) that
\begin{equation}
\label{factor-cn}
M = \frac{|2U(0) - \tilde{U}(0)|}{\max_{x \in \mathbb{R}} |\tilde{U}(x)|} = 2,
\end{equation}
which is the double magnification factor of the rogue wave on the ${\rm cn}$-periodic wave (\ref{red-2})
obtained in \cite{CPnls}.

\subsection{Periodic standing waves with nontrivial phase}

\begin{figure}[h!]
	\centering
	\includegraphics[scale=0.5]{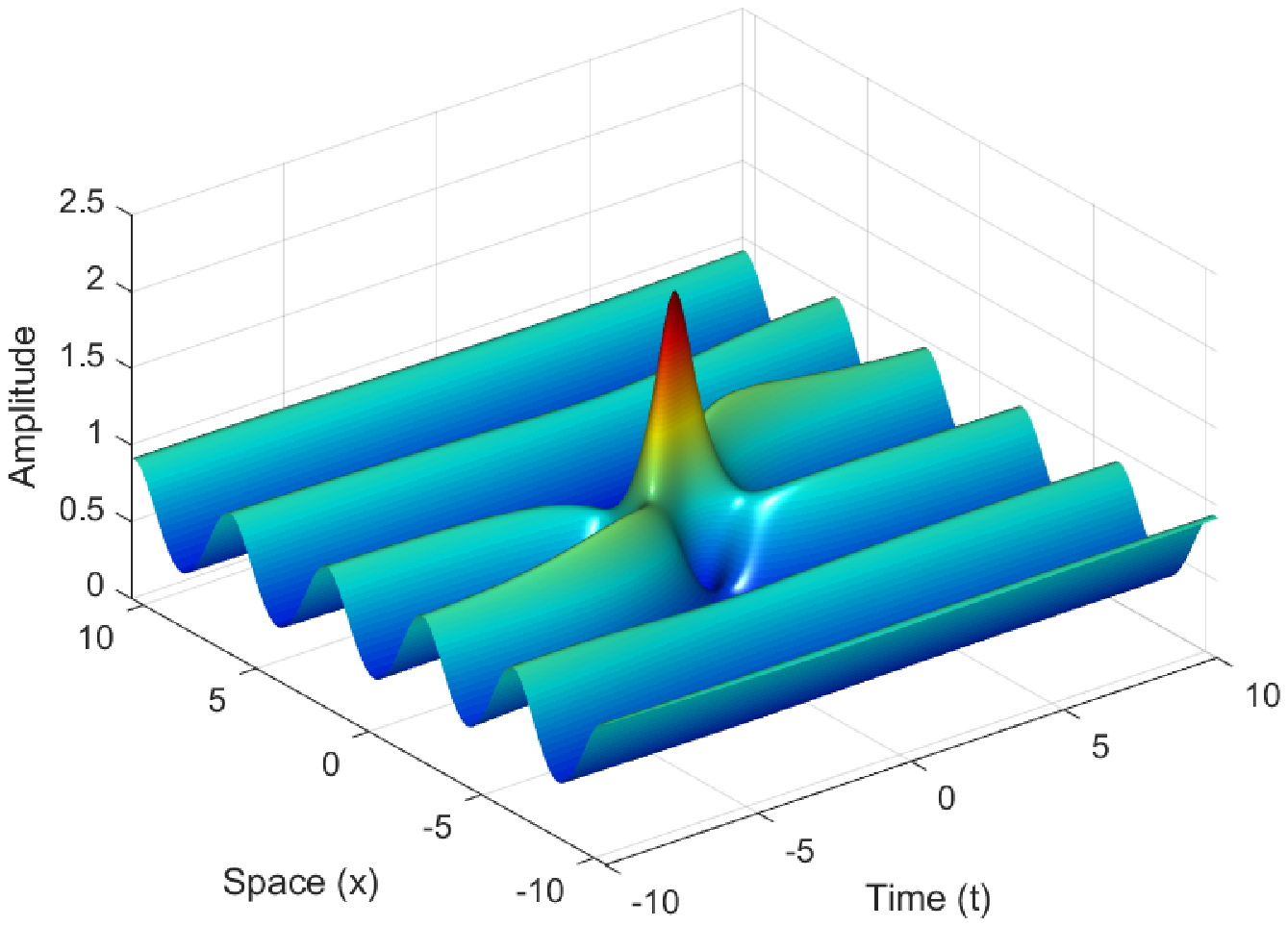}
	\includegraphics[scale=0.5]{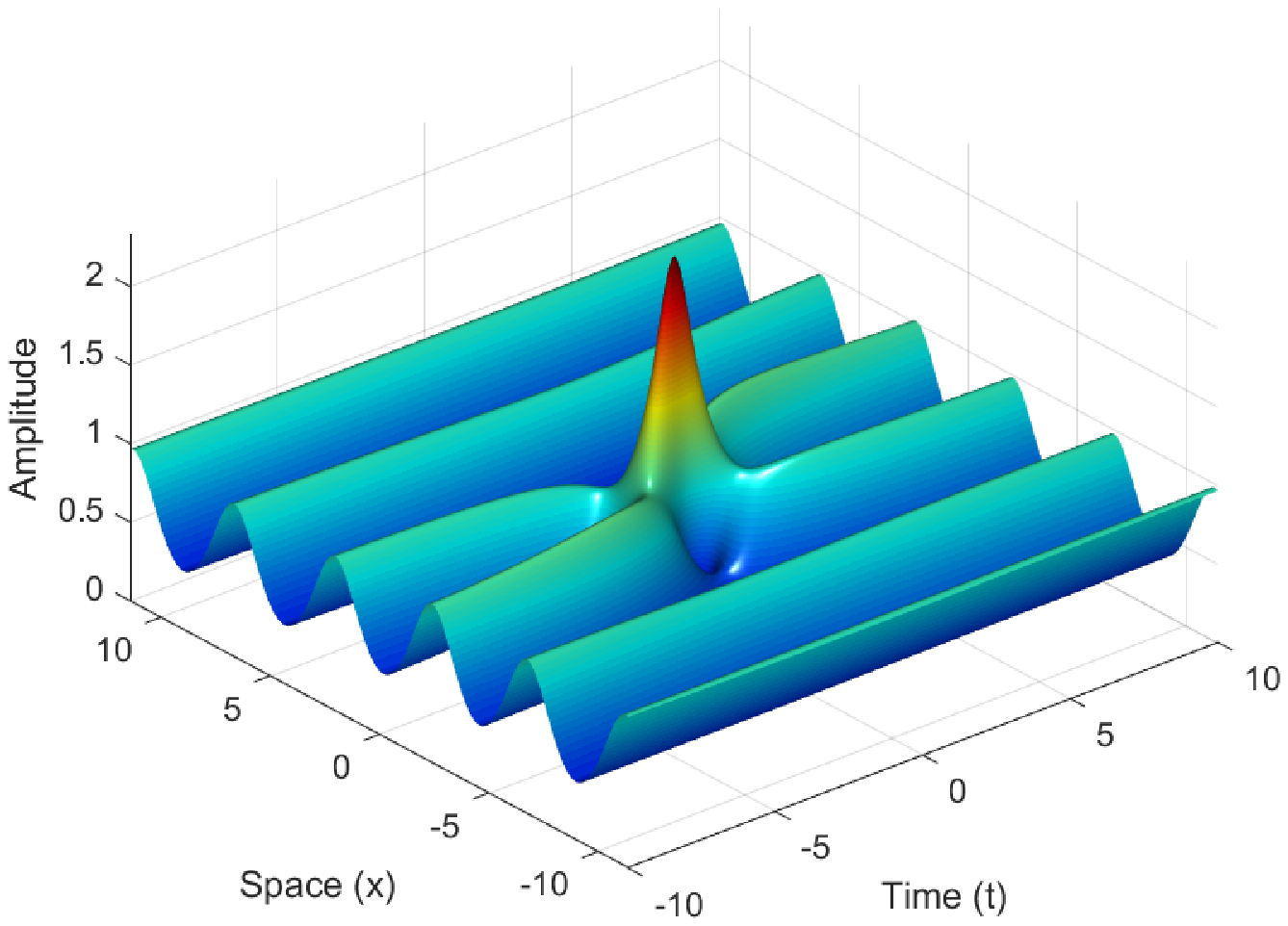}\\
	\includegraphics[scale=0.5]{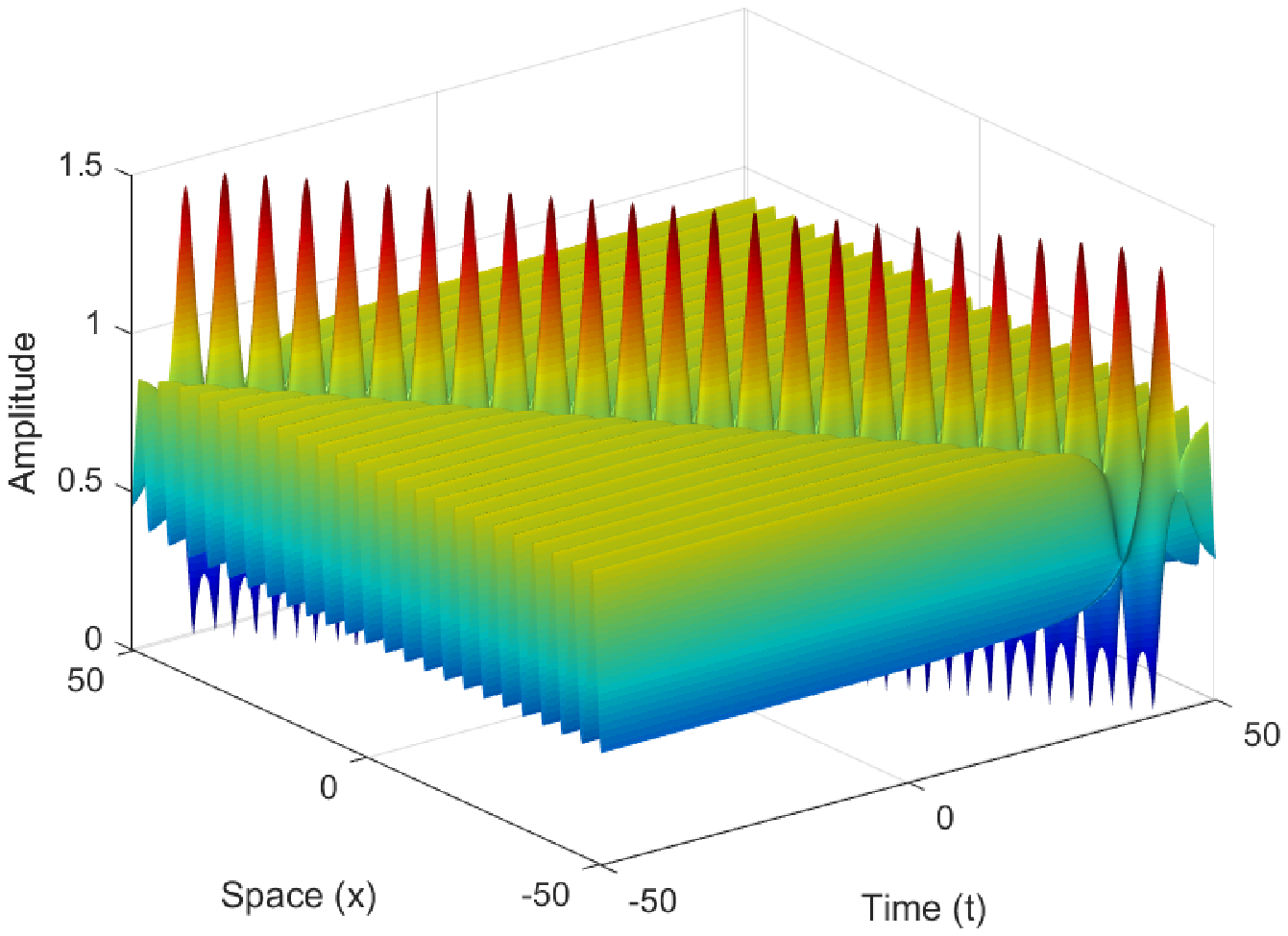}
	\includegraphics[scale=0.5]{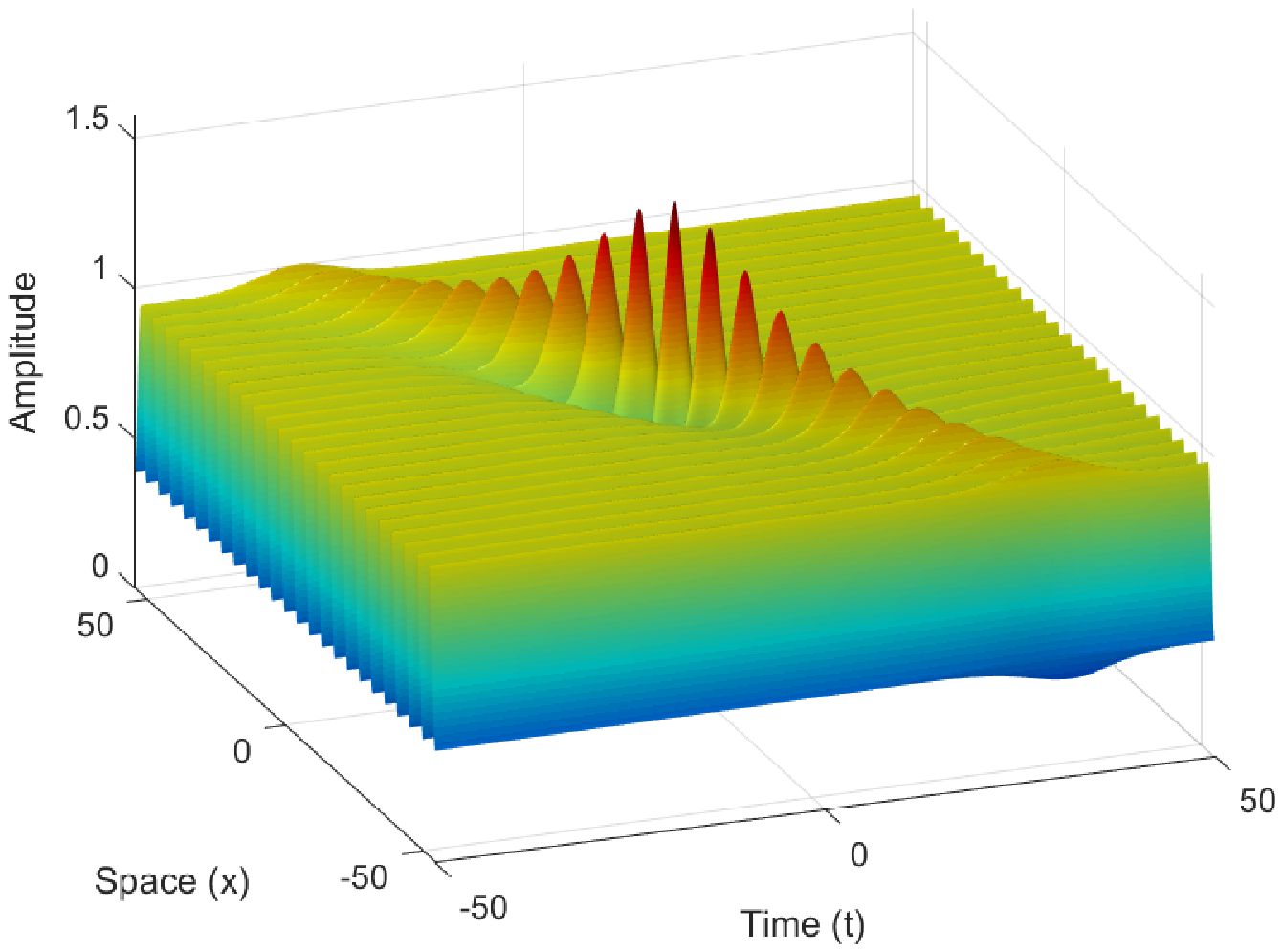}
	\caption{Rogue waves on the periodic wave with non-trivial phase for
    $(\beta,k) = (0.85,0.85)$ (left) and $(\beta,k) = (0.95,0.9)$ (right).
    The upper and lower panels show rogue waves (\ref{rogue-3}) with the upper and lower
    signs respectively.}
	\label{f12}
\end{figure}

For the periodic standing wave given by $U(x) = R(x) e^{i \Theta(x)}$,
we obtain from (\ref{separation-U}), (\ref{separation}), (\ref{separation-3-new}),
(\ref{dn-relation-1}), (\ref{dn-relation-2}), and (\ref{1-fold-rogue-explicit}):
\begin{equation}
\label{rogue-3}
\hat{u} = \left[ \mp \tilde{R} + \frac{4 (\sqrt{\rho_1} \pm \sqrt{\rho}_2)
(\tilde{P}_1^2 \phi_1 - \overline{\tilde{Q}}_1^2 \overline{\phi}_1) + 8(R \pm \tilde{R})}{4 + |\phi_1|^2 {\rm dn}^2(x;k)}
\right] e^{i \Theta} e^{2 i b t},
\end{equation}
where
$$
\tilde{R} =  \frac{\sqrt{\rho_1 \rho_2}}{R(x)} - \frac{i \sqrt{-\rho_3}}{{\rm dn}^2(x;k)} \frac{dR}{dx}
$$
and $\phi_1(x,t)$ is given by (\ref{phi-3}). Fig. \ref{f12} shows rogue waves (\ref{rogue-3}) for the upper sign (upper panels) and
the lower sign (lower panels). The left and right panels correspond to two choices of
$k$ with qualitatively different Lax spectrum on Fig. \ref{f4}. The upper and lower signs
correspond to two choices of $\lambda_1$ in (\ref{lambda-3}).

Since $|\phi_1(x,t)| \to \infty$ as $|x| + |t| \to \infty$
everywhere on the $(x,t)$ plane with the only exception for parameters satisfying the constraint (\ref{boundary}),
we have $\hat{u}(x,t) \sim \mp \tilde{R}(x) e^{i \Theta(x)} e^{2ibt}$ as $|x| + |t| \to \infty$.
It follows from (\ref{Jacobi-explicit-rho}) that
\begin{eqnarray*}
|\tilde{R}(x)|^2 & = & \frac{\rho_1 \rho_2}{\rho(x)} - \frac{\rho_3}{4 \rho(x) {\rm dn}^4(x;k)} \left( \frac{d \rho}{dx} \right)^2\\
& = & \frac{\beta (\beta - k^2) {\rm dn}^2(x;k) + (1-\beta) k^4 {\rm sn}^2(x;k) {\rm cn}^2(x;k)}{(\beta - k^2 {\rm sn}^2(x;k)) {\rm dn}^2(x;k)} \\
& = & \beta - \frac{k^2 {\rm cn}^2(x;k)}{{\rm dn}^2(x;k)} \\
& = & \beta - k^2 {\rm sn}^2(x+K(k);k),
\end{eqnarray*}
where the last equality follows from Table 22.4.3 in \cite{Olver}.
Hence $|\tilde{R}(x)|$ is a translation of $R(x)$ in $x$ by a half period.
On the other hand, since $\phi_1(0,0) = 0$, $R(0) = \sqrt{\rho_1}$, and $\frac{dR}{dx} |_{x=0} = 0$,
and the maximum of $|\hat{u}(x,t)|$ occurs at the origin $(x,t) = (0,0)$ (see Fig. \ref{f12}),
it follows from (\ref{factor-magnification-explicit}) that
\begin{equation}
\label{factor-general}
M = \frac{|2R(0) \pm \tilde{R}(0)|}{\max_{x \in \mathbb{R}} |\tilde{R}(x)|} = 2 \pm \sqrt{1 - \frac{k^2}{\beta}}.
\end{equation}
When $\beta = 1$, $M = 2 \pm \sqrt{1 - k^2}$ coincides with the magnification factor (\ref{factor-dn}) of the ${\rm dn}$-periodic wave (\ref{red-1}).
When $\beta = k^2$, $M = 2$ coincides with the double magnification factor (\ref{factor-cn})
of the ${\rm cn}$-periodic wave (\ref{red-2}).

All the solutions on Fig. \ref{f12} represent the rogue waves on the background of the periodic standing wave
with the exception of the left lower panel, which corresponds to the point $(\beta,k) = (0.85,0.85)$
near the red curve of Fig. \ref{figure-curve} given by the implicit equation (\ref{boundary}).
In this exceptional case, $|\phi_1(x,t)|$ remains bounded along the family of straight lines
in the $(x,t)$ plane given by (\ref{straight-line}). As a result, instead of a rogue wave localized on
the background of the periodic standing wave, the solution in the exceptional case represents
an algebraic soliton propagating on the background of the periodic standing wave. This solution is
similar to the algebraic solitons propagating on the modulationally stable ${\rm dn}$-periodic
waves of the modified KdV equation \cite{CPkdv}. Note
that the algebraic soliton on the periodic standing wave background does not satisfy the limits (\ref{rogue-wave-def}).

\section{Relation to the modulation instability of the periodic waves}

Here we solve the linear equations (\ref{3.1})--(\ref{3.2})
in the case of the periodic standing wave (\ref{separation-U}).
The spectral parameter $\lambda$ is defined in the Lax spectrum of the spectral
problem (\ref{3.1}). Separating variables by
\begin{equation}
\label{lax-eq}
u(x,t) = U(x) e^{2ibt}, \quad \varphi_1(x,t) = \chi_1(x) e^{ibt + t \Omega}, \quad
\varphi_2(x,t) = \chi_2(x) e^{-ibt + t \Omega},
\end{equation}
where $\Omega \in \mathbb{C}$ is another spectral parameter, yields
two spectral problems
\begin{equation}
\label{lin-alg}
\chi_x = \left(\begin{array}{cc} \lambda & U \\ -\bar{U} & -\lambda \end{array} \right) \chi, \quad
\Omega \chi = i \left(\begin{array}{cc}
\lambda^2 + \frac{1}{2} |U|^2 - b & \frac{1}{2} \frac{dU}{dx} + \lambda U \\
\frac{1}{2} \frac{d \bar{U}}{dx} - \lambda\bar{U} & -\lambda^2 - \frac{1}{2} |U|^2 + b \end{array} \right) \chi.
\end{equation}
Since the second spectral problem in (\ref{lin-alg}) is a linear algebraic system, it admits a nonzero solution if and only if
the determinant of the coefficient matrix is zero. The latter condition yields
the $x$-independent relation between $\Omega$ and $\lambda$ in the form
$\Omega^2 + P(\lambda) = 0$, where $P(\lambda)$ is given by (\ref{Polynomial-2}) with $c = 0$
and the first-order invariants (\ref{Constant-2b}) and (\ref{Constant-2a}) with $c = 0$ have been used.

By Theorem 5.1 in \cite{DS}, if $\lambda$ is in the Lax spectrum
of the first spectral problem in (\ref{lin-alg}), then the
squared eigenfunctions $\chi_1^2$ and $\chi_2^2$ determine eigenfunctions
of the linearized NLS equation at the periodic standing wave with the eigenvalues given by
\begin{equation}
\Gamma = 2 \Omega = \pm 2i \sqrt{P(\lambda)}.
\label{eig-stability}
\end{equation}
If ${\rm Re}(\Gamma) > 0$ for $\lambda$ in the Lax spectrum, the periodic standing wave is
called {\em spectrally unstable}.
If the instability band intersects the origin in the $\Gamma$-plane, the periodic standing wave
is called {\em modulationally unstable}.

The Lax spectrum for $\lambda$ in the first spectral problem in
(\ref{lin-alg}) is found with the help of the Floquet--Bloch decomposition in Appendix \ref{appendix-a}.
By Theorem 7.1 in \cite{DS}, $i \mathbb{R}$ belongs to the Lax spectrum
and it follows from (\ref{P-poly-0}) and (\ref{P-poly}) that
$$
P(\lambda) > 0 \quad \mbox{for every } \;\; \lambda \in i \mathbb{R},
$$
hence $\Gamma \in i \mathbb{R}$ in (\ref{eig-stability}) for every $\lambda \in i \mathbb{R}$.
Therefore, the spectral instability of the periodic standing wave arises only if $\lambda$ belongs to
the finite band(s) with ${\rm Re}(\lambda) \neq 0$, see Figs. \ref{f1}, \ref{f2}, and \ref{f4}.
Recently, this conclusion was generalized for other nonlinear integrable equations in \cite{DU2}.

For the ${\rm dn}$-periodic wave (\ref{red-1}) it follows from (\ref{P-poly-0})
that $P(\lambda) < 0$ for every $|\lambda| \in (\lambda_2^+,\lambda_1^+)$ and $P(\lambda_{1,2}^+) = 0$,
where $\lambda_{1,2}^+$ are given by (\ref{eig-dn}). The corresponding values of $\Gamma$ in (\ref{eig-stability})
belongs to the finite segment on the real line which touches the origin since $\Gamma = 0$ if
$\lambda = \lambda_{1,2}^+$. Hence, the ${\rm dn}$-periodic wave (\ref{red-1})
is modulationally unstable and the rogue waves constructed for $\lambda = \lambda_{1,2}^+$ on Fig. \ref{f9}
correspond to the end points of the modulation instability band.

\begin{figure}[h!]
	\centering
	\includegraphics[scale=0.5]{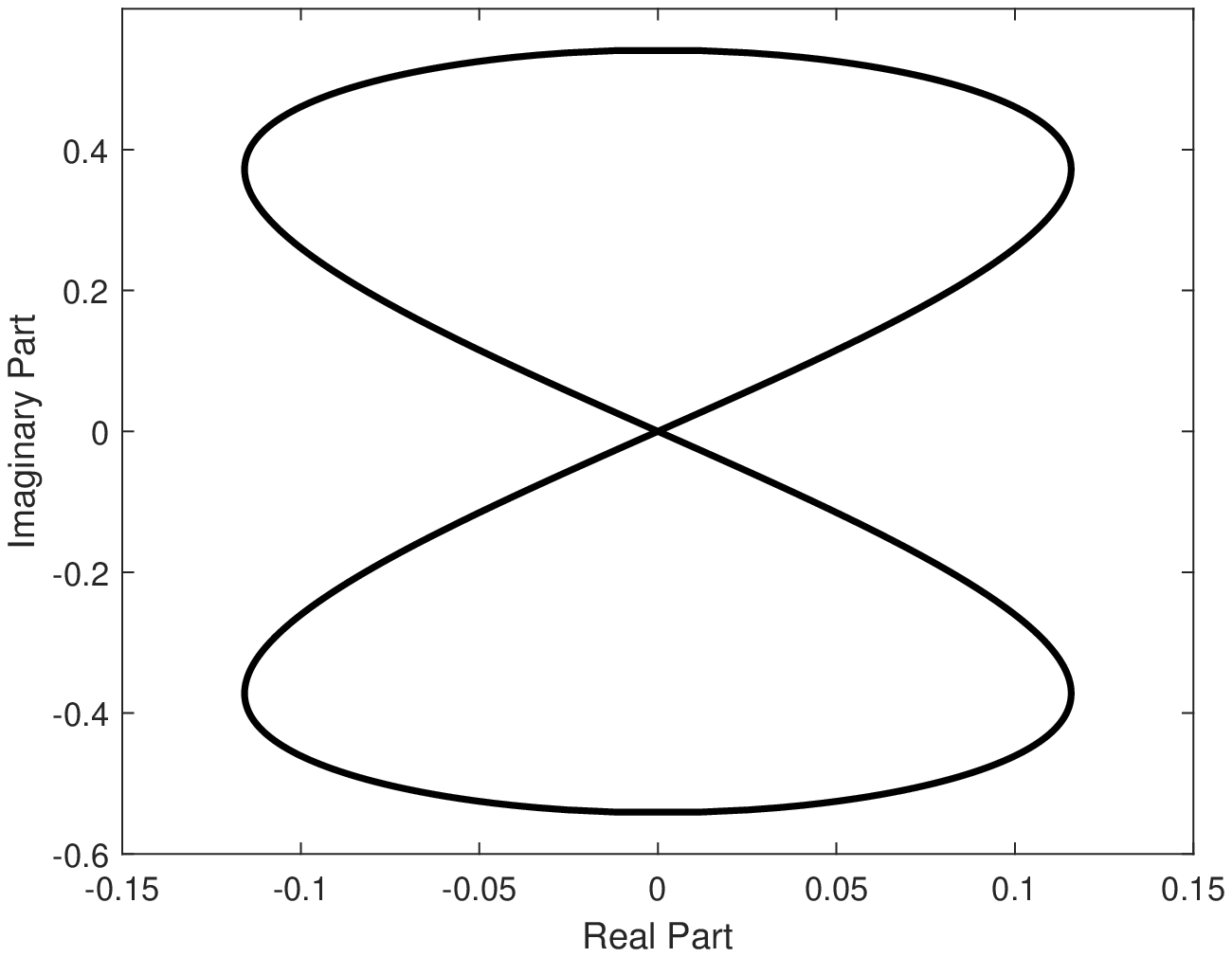}
	\includegraphics[scale=0.5]{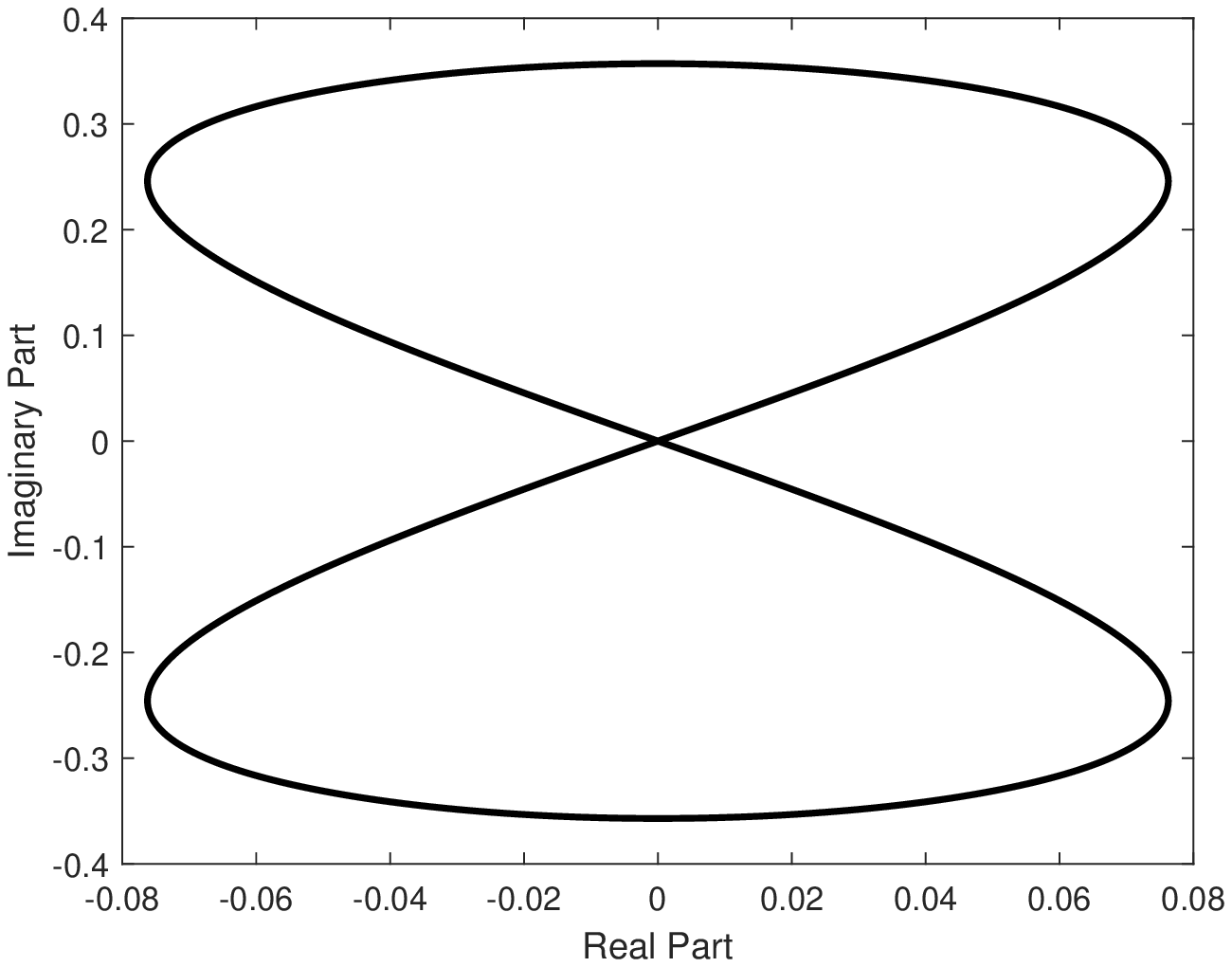}
	\caption{Modulation instability bands for the ${\rm cn}$-periodic wave (\ref{red-2})
    with $k=0.85$ (left) and $k = 0.95$ (right).}
	\label{f17}
\end{figure}

Similarly, for the ${\rm cn}$-periodic wave (\ref{red-2}),
the trace of $\Gamma$ in (\ref{eig-stability}) on the complex plane
is shown on Fig. \ref{f17}. The curves are obtained when
$\lambda$ changes along the two bands of the Lax spectrum in Fig. \ref{f2} with
${\rm Re}(\lambda) \neq 0$. Note that each curve on Fig. \ref{f17} is covered twice.
The modulation instability bands on Fig. \ref{f17} are similar for both examples
of the ${\rm cn}$-periodic wave (\ref{red-2}) with two different Lax spectrum on Fig. \ref{f2}.
Again, $\Gamma = 0$ at $\lambda = \lambda_{1,2}^+$, hence
the rogue waves constructed for $\lambda = \lambda_{1,2}^+$ on Fig. \ref{f10}
correspond to the end points of the modulation instability band.

\begin{figure}[h!]
	\centering
	\includegraphics[scale=0.5]{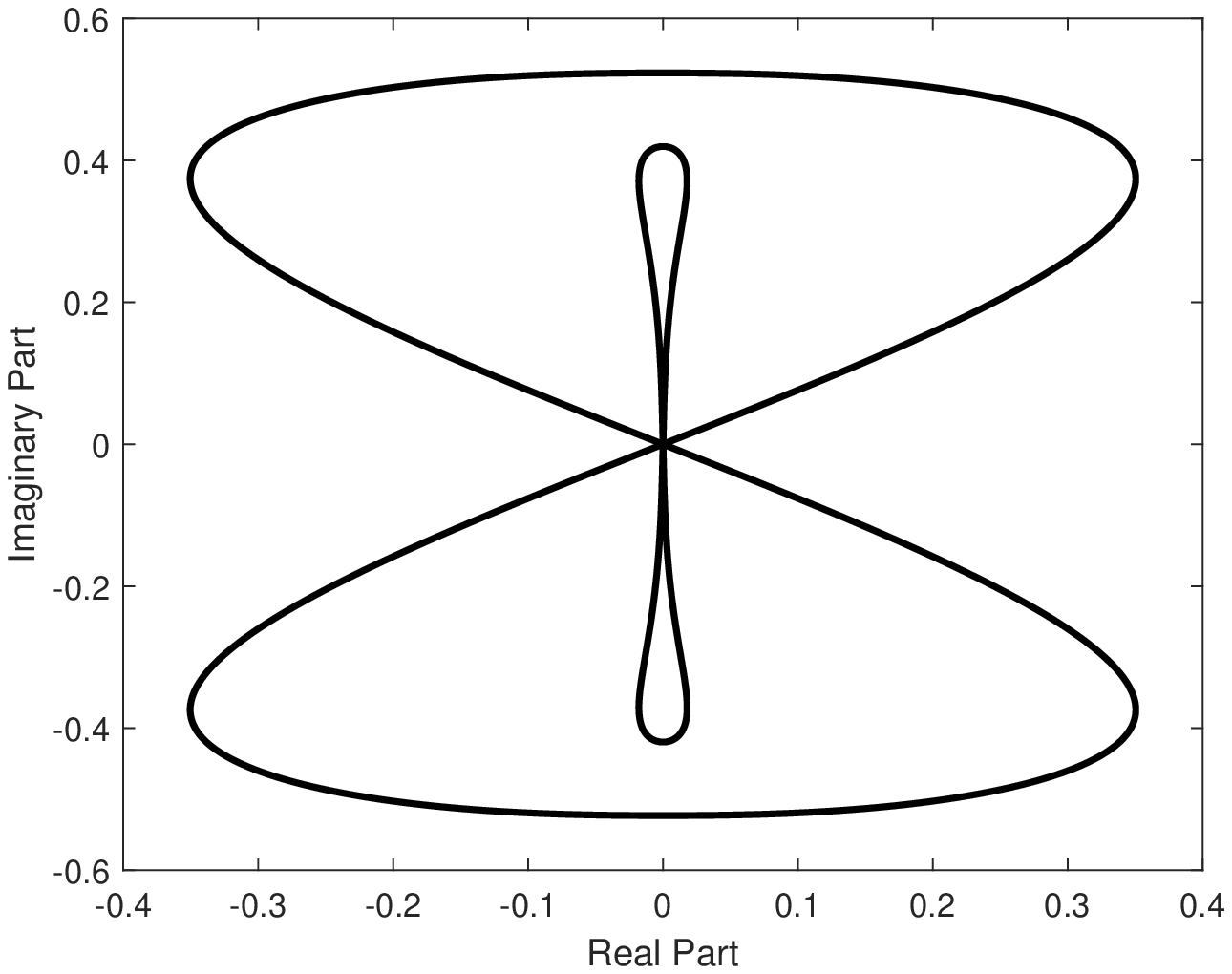}
	\includegraphics[scale=0.5]{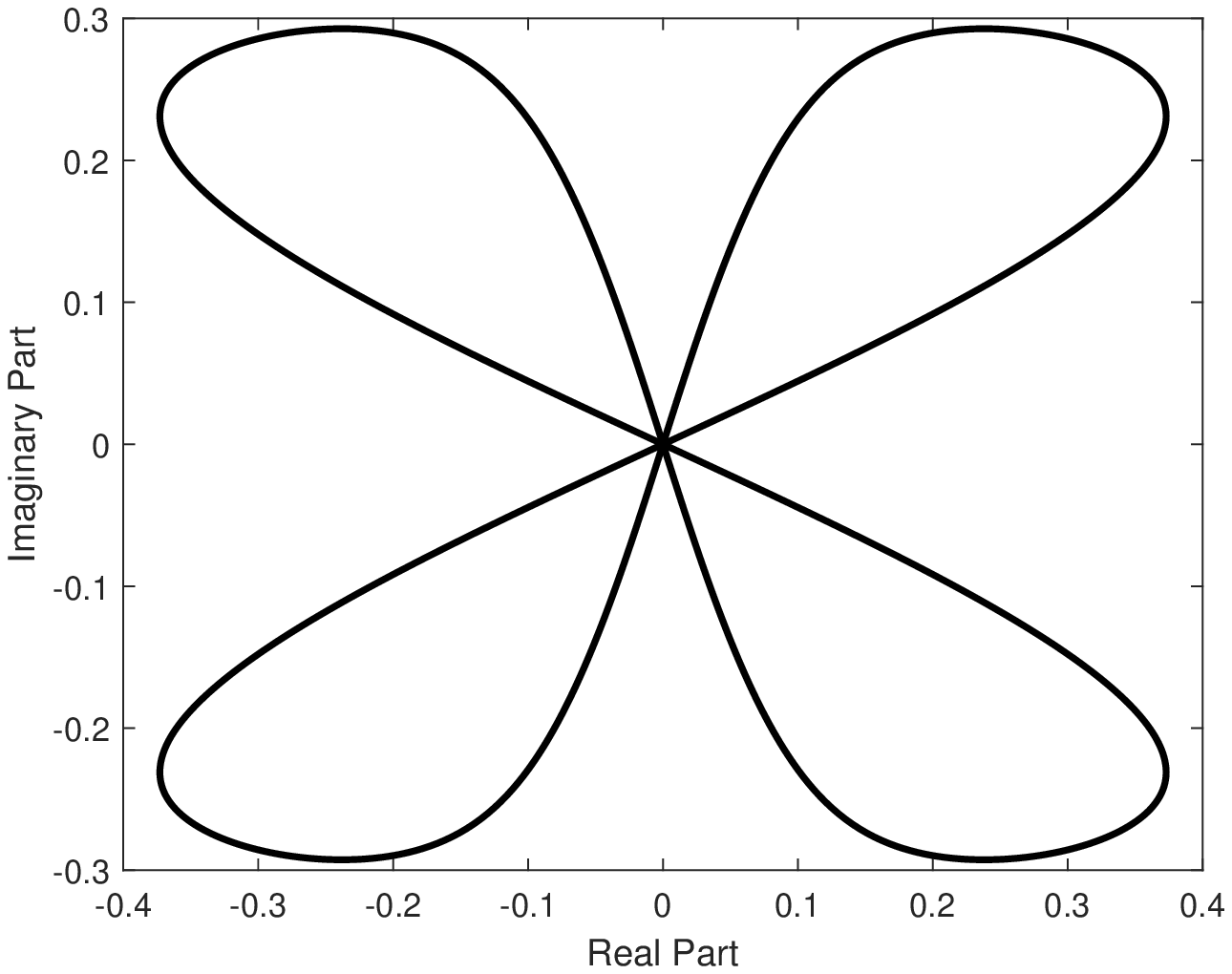}
	\caption{Modulation instability bands for the periodic wave (\ref{Jacobi-explicit-rho})
    with $(\beta,k) = (0.85,0.85)$ (left) and $(\beta,k) = (0.95,0.9)$ (right).}
	\label{f15}
\end{figure}

For the periodic standing wave with non-trivial phase,
the trace of $\Gamma$ in (\ref{eig-stability}) is shown on Fig. \ref{f15}
when $\lambda$ changes along the two bands of the Lax spectrum on Fig. \ref{f4} with ${\rm Re}(\lambda) \neq 0$.
The symmetry of the Lax spectrum on Fig. \ref{f4} is broken and each curve
on Fig. \ref{f15} is covered once. It follows from Fig. \ref{figure-curve} that the point $(\beta,k) = (0.85,0.85)$
is selected near the boundary (\ref{boundary}). This boundary coincides with the condition
under which the second band of the modulation instability is
tangential to the imaginary axis of $\Gamma$ at $\Gamma = 0$
as on the left panel of Fig. \ref{f15}. Indeed, the nonlinear equation  (\ref{boundary})
was found in equation (106) in \cite{DS} from the above condition.

It follows from different types of the rogue waves in Fig. \ref{f12} that the rogue wave
on the background of the periodic standing wave satisfying the limits (\ref{rogue-wave-def}) exists if the modulation
instability band is transverse to the imaginary axis at $\Gamma = 0$, whereas the algebraic soliton on the periodic standing wave
exists if the modulation instability band is tangential to the imaginary axis at $\Gamma = 0$.
We emphasize again that the algebraic soliton on the periodic standing wave does not satisfy the limits (\ref{rogue-wave-def}).

We conclude the paper by reiterating the question on how to define the parameter $\lambda_1$
in the one-fold Darboux transformation (\ref{1-fold}) in order to generate the
rogue waves on the background of the periodic standing wave satisfying the limits (\ref{rogue-wave-def}).
If $\lambda_1 \in i \mathbb{R}$, then $\hat{u} = u$ and no new solution is obtained.
If $\lambda_1 \notin i \mathbb{R}$ is outside the other bands of the Lax spectrum, then the one-fold transformation
(\ref{1-fold}) generates the recurrent pattern of rogue waves. In the case of the constant-amplitude
waves, such recurrent rogue waves are usually referred to as the Kuznetsov--Ma breathers.
These rogue waves do not satisfy the limits (\ref{rogue-wave-def}).

On the other hand, if $\lambda_1 \notin i \mathbb{R}$ is inside the other bands of the Lax spectrum,
then the one-fold transformation (\ref{1-fold}) generates a periodic perturbation
on the periodic wave background which grows and decays exponentially in time
due to modulation instability with the growth rate $\Gamma$ in (\ref{eig-stability}).
In the context of the constant-amplitude wave, the space-periodic and time-localized solutions are usually referred to
as the Akhmediev breathers. Since the perturbation period is different from the period
of the periodic wave background, such solutions are generally quasi-periodic in space and exponentially
localized in time. These rogue waves also satisfy the limits (\ref{rogue-wave-def}) but do not
represent isolated rogue waves.

Isolated rogue waves on the background of the periodic standing wave are
generated by picking the value of $\lambda_1$ exactly at the end points
of the bands of the Lax spectrum outside $i \mathbb{R}$. Isolated rogue waves
satisfy the limits (\ref{rogue-wave-def}) with the only exception when
$|\phi_1(x,t)|$ remains bounded along a family of the straight lines in the $(x,t)$-plane,
as it happens for the periodic standing wave with parameters satisfying the condition (\ref{boundary}).
This exception which generates an algebraic soliton on the periodic standing wave
corresponds to the case when the modulation instability band is tangential
to the imaginary axis at $\Gamma = 0$.

The precise values of $\lambda_1$
at the end points of the Lax spectrum are captured by the algebraic method with one eigenvalue
developed in this article. This method complements the previous characterization
of the end points of the bands of the Lax spectrum outside $i \mathbb{R}$
with the resolvent method in \cite{Kam,Kam-review}.

\vspace{0.25cm}

{\bf Acknowledgement.} Analytical work on this project was supported by the National
Natural Science Foundation of China (No. 11971103). Numerical work was supported by
the Russian Science Foundation  (No.19-12-00253).

\appendix
\section{Floquet--Bloch decomposition of the Lax spectrum}
\label{appendix-a}

If the entries of the matrix $Q$ in the linear equation \eqref{3.1} are periodic in $x$ with
the same period $L$, then Floquet's Theorem guarantees that bounded solutions
of the linear equation (\ref{3.1}) can be represented in the form:
\begin{equation}\label{A1FB}
\varphi(x) = \left(\begin{array}{cc} \breve{p}_{1}(x)  \\ \breve{q}_{1}(x)
\end{array} \right)e^{i \mu x},
\end{equation}
where $\breve{p}_{1}(x) = \breve{p}_{1}(x + L)$ , $\breve{q}_{1}(x) =
\breve{q}_{1}(x + L)$, and $\mu \in \left[-\frac{\pi}{L},\frac{\pi}{L} \right]$.
The Lax spectrum in the linear equation (\ref{3.1}) is formed by all admissible
values of $\lambda$, for which the solutions are bounded in the form (\ref{A1FB}),
where $i \mu$ is referred to as the Floquet exponent. When $\mu=0$ and $\mu = \pm \frac{\pi}{L}$, the
solutions (\ref{A1FB}) are periodic and anti-periodic, respectively.

Substituting \eqref{A1FB} into the linear equation \eqref{3.1} and re-arranging
the terms yields the eigenvalue problem:
\begin{equation}\label{A12}
\left(\begin{array}{cc} \frac{d}{dx} + i \mu & -u \\ -\bar{u} & -\frac{d}{dx} - i \mu
\end{array} \right) \left(\begin{array}{cc} \breve{p}_{1}\\ \breve{q}_{1}
\end{array} \right)= \lambda \left(\begin{array}{cc} \breve{p}_{1} \\
\breve{q}_{1}
\end{array} \right),
\end{equation}
for which we are looking for $L$-periodic solutions $(\breve{p}_{1},\breve{q}_{1})$
at a discrete set of admissible values of $\lambda_1$.

The numerical scheme of computing the eigenvalues $\lambda$ is based on the discretization
of the interval $[0,L]$ with $N+1$ equally spaced grid points and using the highly accurate
central difference approximation  of derivatives (up to the 12th order of accuracy). MATLAB's
eigenvalue solver is used to compute all eigenvalues $\lambda$ in the discretization
of the eigenvalue problem \eqref{A12}. Tracing the set of eigenvalues $\lambda$ for $\mu \in \left[-\frac{\pi}{L},\frac{\pi}{L} \right]$
gives the band of the Lax spectrum in the $\lambda$ plane shown on Figs. \ref{f1}, \ref{f2}, and \ref{f4}.

\section{Proof of identities (\ref{dn-relation-1}) and (\ref{dn-relation-2})}
\label{appendix-b}

We substitute (\ref{separation-3-new}) into (\ref{pq-3}),
recall that $R^2 \frac{d \Theta}{dx} = -2a = \sqrt{\rho_1 \rho_2} \sqrt{-\rho_3}$, and obtain
\begin{eqnarray}
\left\{ \begin{array}{l}
\tilde{P}_1^2 = \frac{1}{2(\sqrt{\rho_1} \pm \sqrt{\rho_2})} \left( \frac{dR}{dx} + (\sqrt{\rho_1} \pm \sqrt{\rho_2}) R
+ i\sqrt{-\rho_3} (\pm R + \sqrt{\rho_1 \rho_2} R^{-1}) \right), \\
\tilde{Q}_1^2 = \frac{1}{2(\sqrt{\rho_1} \pm \sqrt{\rho_2})} \left( -\frac{dR}{dx} + (\sqrt{\rho_1} \pm \sqrt{\rho_2}) R
+ i\sqrt{-\rho_3} (\pm R + \sqrt{\rho_1 \rho_2} R^{-1}) \right), \\
\tilde{P}_1 \tilde{Q}_1 = -\frac{1}{2(\sqrt{\rho_1} \pm \sqrt{\rho_2})} \left( R^2 \pm \sqrt{\rho_1 \rho_2} \pm i
\sqrt{-\rho_3} (\sqrt{\rho_1} \pm \sqrt{\rho_2})  \right).\end{array} \right.
\label{pq-3-new}
\end{eqnarray}
By using (\ref{R-equation}) with (\ref{roots}) and $\rho = R^2$, we obtain
$(|\tilde{P}_1|^2 + |\tilde{Q}_1|^2)^2 = \rho(x) - \rho_3 = {\rm dn}^2(x;k)$,
where the last identity follows from (\ref{Jacobi-explicit-rho}).
Extracting the square root yields (\ref{dn-relation-1}).

In order to prove (\ref{dn-relation-2}), we compute from (\ref{pq-3-new}):
$$
\tilde{P}_1^2 \overline{\tilde{Q}}_1^2 = \frac{\left[ -\left(\frac{dR}{dx} \right)^2
+ (\sqrt{\rho_1} \pm \sqrt{\rho_2})^2 R^2 -\rho_3 (\pm R^2 + \sqrt{\rho_1 \rho_2} R^{-1})^2
- 2i \sqrt{-\rho_3} (\pm R^2 + \sqrt{\rho_1 \rho_2} R^{-1}) \frac{dR}{dx} \right]}{4(\sqrt{\rho_1} \pm \sqrt{\rho_2})^2}.
$$
By using (\ref{R-equation}) with (\ref{roots}) and (\ref{Jacobi-explicit-rho}),
we check directly that
$$
-\left(\frac{dR}{dx}\right)^2 + (\sqrt{\rho_1} \pm \sqrt{\rho_2})^2 R^2 -\rho_3 (\pm R^2 + \sqrt{\rho_1 \rho_2} R^{-1})^2 =
\frac{{\rm dn}^2(x;k)}{R^2} (\pm R^2 + \sqrt{\rho_1 \rho_2})^2 + \frac{\rho_3}{{\rm dn}^2(x;k)} \left( \frac{dR}{dx} \right)^2,
$$
which yields
$$
\tilde{P}_1^2 \overline{\tilde{Q}}_1^2 = \frac{1}{4(\sqrt{\rho}_1 \pm \sqrt{\rho}_2)^2}
\left[ \frac{{\rm dn}(x;k)}{R} ( \pm R^2 + \sqrt{\rho_1 \rho_2}) - \frac{i \sqrt{-\rho_3}}{{\rm dn}(x;k)} \frac{dR}{dx} \right]^2.
$$
Extracting the square root and picking the negative sign by using the limiting expression (\ref{cn-identity-2})
for $\rho_1 = k^2$, $\rho_2 = 0$, and $\rho_3 = -(1-k^2)$ yields the expression (\ref{dn-relation-2}).


\begin{thebibliography}{99}


\bibitem{AZ1} D.S. Agafontsev and V.E. Zakharov, ``Integrable turbulence and formation of
rogue waves", Nonlinearity {\bf 28} (2015), 2791--2821.

\bibitem{AZ2} D.S. Agafontsev and V.E. Zakharov, ``Integrable turbulence generated from modulational
instability of cnoidal waves", Nonlinearity {\bf 29} (2016), 3551--3578.

\bibitem{Akh85} N.N. Akhmediev, V.M. Eleonsky, and N.E. Kulagin, ``Generation of periodic trains
of picosecond pulses in an optical fiber: Exact solutions", Sov. Phys. JETP {\bf 62} (1985), 894--899.

\bibitem{Akh} N. Akhmediev, A. Ankiewicz, and J.M. Soto-Crespo,
``Rogue waves and rational solutions of the nonlinear Schr\"{o}dinger equation",
Phys. Rev. E {\bf 80} (2009), 026601 (9 pages).

\bibitem{Tovbis1} M. Bertola, G.A. El, and A. Tovbis, ``Rogue waves in multiphase solutions of the focusing
nonlinear Schr\"{o}dinger equation", Proc. R. Soc. Lond. A {\bf 472} (2016), 20160340 (12 pages).

\bibitem{Tovbis2} M. Bertola and A. Tovbis, ``Maximal amplitudes of finite-gap solutions for
the focusing nonlinear Schr\"{o}dinger equation", Comm. Math. Phys. {\bf 354} (2017), 525--547.

\bibitem{Bil1} D. Bilman and R. Buckingham, ``Large-order asymptotics for multiple-pole solitons of the focusing
nonlinear Schr\"{o}dinger equation", J. Nonlinear Sci. {\bf 29} (2019), 2185--2229.

\bibitem{Bil2} D. Bilman and P. D. Miller, ``A robust inverse scattering transform for the focusing nonlinear
Schr\"{o}dinger equation", Comm. Pure Appl. Math {\bf 72} (2019), 1722--1805.

\bibitem{Bil3} D. Bilman, L. Ling and P.D. Miller, ``Extreme superposition: rogue waves of infinite order
and the Painlev\'{e}-III hierarchy", arXiv:1806.00545 (2018).

\bibitem{Biondini} G. Biondini, S. Li, D. Mantzavinos, and S. Trillo, ``Universal behavior of modulationally unstable media",
SIAM Review {\bf 60} (2018), 888--908.

\bibitem{BHJ} J.C. Bronski, V.M. Hur, and M.A. Johnson, ``Modulational instability in equations of KdV type",
in {\em New approaches to nonlinear waves},  Lecture Notes in Phys. {\bf 908} (Springer, Cham, 2016),
pp. 83--133.

\bibitem{CalSch} A. Calini and C.M. Schober, ``Characterizing JONSWAP rogue waves and their
statistics via inverse spectral data",  Wave Motion {\bf 71} (2017), 5--17.

\bibitem{Cao1} C.W. Cao and X.G. Geng, ``Classical integrable systems generated through nonlinearization of eigenvalue problems",
{\em Nonlinear physics (Shanghai, 1989)}, pp. 68--78 (Research Reports in Physics, Springer, Berlin, 1990).

\bibitem{CPkdv} J. Chen and D.E. Pelinovsky, ``Rogue periodic waves in the modified Korteweg-de Vries equation",
Nonlinearity {\bf 31} (2018), 1955--1980.

\bibitem{CPnls} J. Chen and D.E. Pelinovsky, ``Rogue periodic waves in the focusing nonlinear Schr\"{o}dinger equation",
Proc. R. Soc. Lond. A {\bf 474} (2018), 20170814 (18 pages).

\bibitem{CPW} J. Chen, D.E. Pelinovsky, and R.E. White, ``Rogue waves on the double-periodic background in the
focusing nonlinear Schr\"{o}dinger equation", Phys. Rev. E {\bf 100} (2019), 052219 (18 pages).

\bibitem{DS} B. Deconinck and B.L. Segal, ``The stability spectrum for elliptic
solutions to the focusing NLS equation", Physica D {\bf 346} (2017), 1--19.

\bibitem{DU} B. Deconinck and J. Upsal, ``The orbital stability of elliptic solutions of the focusing nonlinear
Schr\"{o}dinger equation", arXiv: 1901.08702 (2019).

\bibitem{DU2} B. Deconinck and J. Upsal, ``Real Lax spectrum implies spectral stability", arXiv: 1909.10119 (2019).

\bibitem{Matveev} P. Dubard and V.B. Matveev, ``Multi-rogue waves solutions: from the NLS to the KP-I equation",
Nonlinearity {\bf 26}  (2013),  R93--R125.

\bibitem{Feng} B.F. Feng, L. Ling, and D.A. Takahashi, ``Multi-breathers and high order rogue waves for the
nonlinear Schr\"{o}dinger equation on the elliptic function background", Stud. Appl. Math. (2020), in press.

\bibitem{GS1} P.G. Grinevich and P.M. Santini, ``The finite gap method and the analytic description of
the exact rogue wave recurrence in the periodic NLS Cauchy problem", Nonlinearity {\bf 31} (2018), 5258--5308.

\bibitem{GS2} P.G. Grinevich and P.M. Santini, ``The finite gap method and the periodic NLS Cauchy problem
of the anomalous waves, for a finite number of unstable modes", Russ. Math. Surv. {\bf 74} (2019), 211--263.

\bibitem{Kam} A.M. Kamchatnov, ``On improving the effectiveness of periodic solutions of the NLS
and DNLS equations", J. Phys. A: Math. Gen. {\bf 23} (1990), 2945--2960.

\bibitem{Kam-review} A.M. Kamchatnov, ``New approach to periodic solutions of integrable equations and nonlinear
theory of modulational instability", Phys. Rep. {\bf 286} (1997), 199--270.

\bibitem{Kedziora} D.J. Kedziora, A. Ankiewicz, and N. Akhmediev, ``Rogue waves and solitons on a cnoidal
background", Eur. Phys. J. Special Topics {\bf 223} (2014), 43--62.

\bibitem{JYang} Y. Ohta and J. Yang, ``General high-order rogue waves and their
dynamics in the nonlinear Schr\"{o}dinger equation", Proc. R. Soc. Lond. A {\bf 468} (2012), 1716--1740.

\bibitem{Olver} F.W.J. Olver, D.W. Lozier, R.F. Boisvert, and C.W. Clark, ``NIST Handbook of Mathematical Functions'', Cambridge University Press, ISBN: 978-0-521-19225-5, (2010).

\bibitem{Peregrine} D.H. Peregrine, ``Water waves, nonlinear Schr\"{o}dinger equations and their solutions",
J. Austral. Math. Soc. B. {\bf 25} (1983), 16--43.

\bibitem{ZakOst} V.E. Zakharov and L.A. Ostrovsky, ``Modulation
instability: The beginning", Physica D {\bf 238} (2009), 540--548.

\bibitem{ZhouJMP} R.G. Zhou, ``Nonlinearization of spectral problems of the
nonlinear Schr\"{o}dinger equation and the real-valued modified Korteweg de
Vries equation", J. Math. Phys. {\bf 48} (2007), 013510 (9 pages).

\bibitem{ZhouStudies} R.G. Zhou, ``Finite-dimensional integrable Hamiltonian systems
related to the nonlinear Schr\"{o}dinger equation", Stud. Appl. Math. {\bf 123} (2009), 311--335.
\end{thebibliography}
\end{document}